\documentclass[aps,prx,superscriptaddress,amsfonts,amsmath,amssymb,reprint,showpacs,showkeys,floatfix]{revtex4-1}

\usepackage{graphicx}
\usepackage{amsmath}
\usepackage{amstext}
\usepackage{amsfonts}
\usepackage{amssymb}
\usepackage{amsbsy}
\usepackage{verbatim}
\usepackage{placeins}
\usepackage{helvet}
\usepackage{dcolumn}
\usepackage{nicefrac}
\usepackage{textcomp}
\usepackage{color}
\usepackage{url}
\usepackage{bm}
\usepackage[colorlinks=true, urlcolor=blue, linkcolor=blue, citecolor=blue, pdftex]{hyperref}
\usepackage{gensymb}
\usepackage{pdfpages}

\begin{document}

\title{Paramagnetism in the kagome compounds (Zn,Mg,Cd)Cu$_{3}$(OH)$_{6}$Cl$
_{2}$}
\author{Yasir Iqbal}
\email{yiqbal@physik.uni-wuerzburg.de}
\affiliation{Institute for Theoretical Physics and Astrophysics, Julius-Maximilian's
University of W\"urzburg, Am Hubland, D-97074 W\"urzburg, Germany}
\author{Harald O. Jeschke}
\email{jeschke@itp.uni-frankfurt.de}
\affiliation{Institut f\"ur Theoretische Physik, Goethe-Universit{\"a}t Frankfurt,
Max-von-Laue-Stra{\ss }e 1, D-60438 Frankfurt am Main, Germany}
\author{Johannes Reuther}
\email{reuther@zedat.fu-berlin.de}
\affiliation{Dahlem Center for Complex Quantum Systems and Fachbereich Physik, Freie Universit{\"a}t Berlin, D-14195 Berlin, Germany}
\affiliation{Helmholtz-Zentrum Berlin f{\"u}r Materialien und Energie, D-14109 Berlin, Germany}
\author{Roser Valent\'i}
\email{valenti@itp.uni-frankfurt.de}
\affiliation{Institut f\"ur Theoretische Physik, Goethe-Universit{\"a}t Frankfurt,
Max-von-Laue-Stra{\ss }e 1, D-60438 Frankfurt am Main, Germany}
\author{I. I. Mazin}
\email{mazin@nrl.navy.mil}
\affiliation{Center for Computational Materials Science, Naval Research Laboratory, Code
6390, 4555 Overlook Ave, SW, Washington, DC 20375, United States of America}
\author{Martin Greiter}
\email{greiter@physik.uni-wuerzburg.de}
\affiliation{Institute for Theoretical Physics and Astrophysics, Julius-Maximilian's
University of W\"urzburg, Am Hubland, D-97074 W\"urzburg, Germany}
\author{Ronny Thomale}
\email{rthomale@physik.uni-wuerzburg.de}
\affiliation{Institute for Theoretical Physics and Astrophysics, Julius-Maximilian's
University of W\"urzburg, Am Hubland, D-97074 W\"urzburg, Germany}
\date{\today}

\begin{abstract}
Frustrated magnetism on the kagome lattice has been a  fertile ground for
rich and fascinating physics, ranging from  experimental evidence of a spin
liquid to theoretical predictions  of exotic superconductivity. Among
experimentally realized  spin-$\frac{1}{2}$ kagome magnets, herbertsmithite,
kapellasite,  and haydeeite [(Zn,Mg)Cu$_{3}$(OH)$_{6}$Cl$_{2}$] are all
well  described by a three-parameter Heisenberg model, but they exhibit 
distinctly different physics. We address the problem using a 
pseudofermion functional renormalization-group  approach and analyze the
low-energy physics in the experimentally accessible parameter range. Our
analysis places kapellasite and  haydeeite near the boundaries between
magnetically ordered and  disordered phases, implying that slight
modifications could  dramatically affect their magnetic properties. Inspired
by this,  we perform \textit{ab initio} density functional theory calculations of  (Zn,Mg,Cd)Cu$_{3}$
(OH)$_{6}$Cl$_{2}$ at various pressures. Our  results suggest that by
varying pressure and composition one can  traverse a paramagnetic regime
between different magnetically ordered  phases.
\end{abstract}

\pacs{75.10.Jm, 75.10.Kt, 75.40.Mg, 05.10.Cc}

\maketitle

\begin{figure}
\includegraphics[width=1.0\columnwidth]{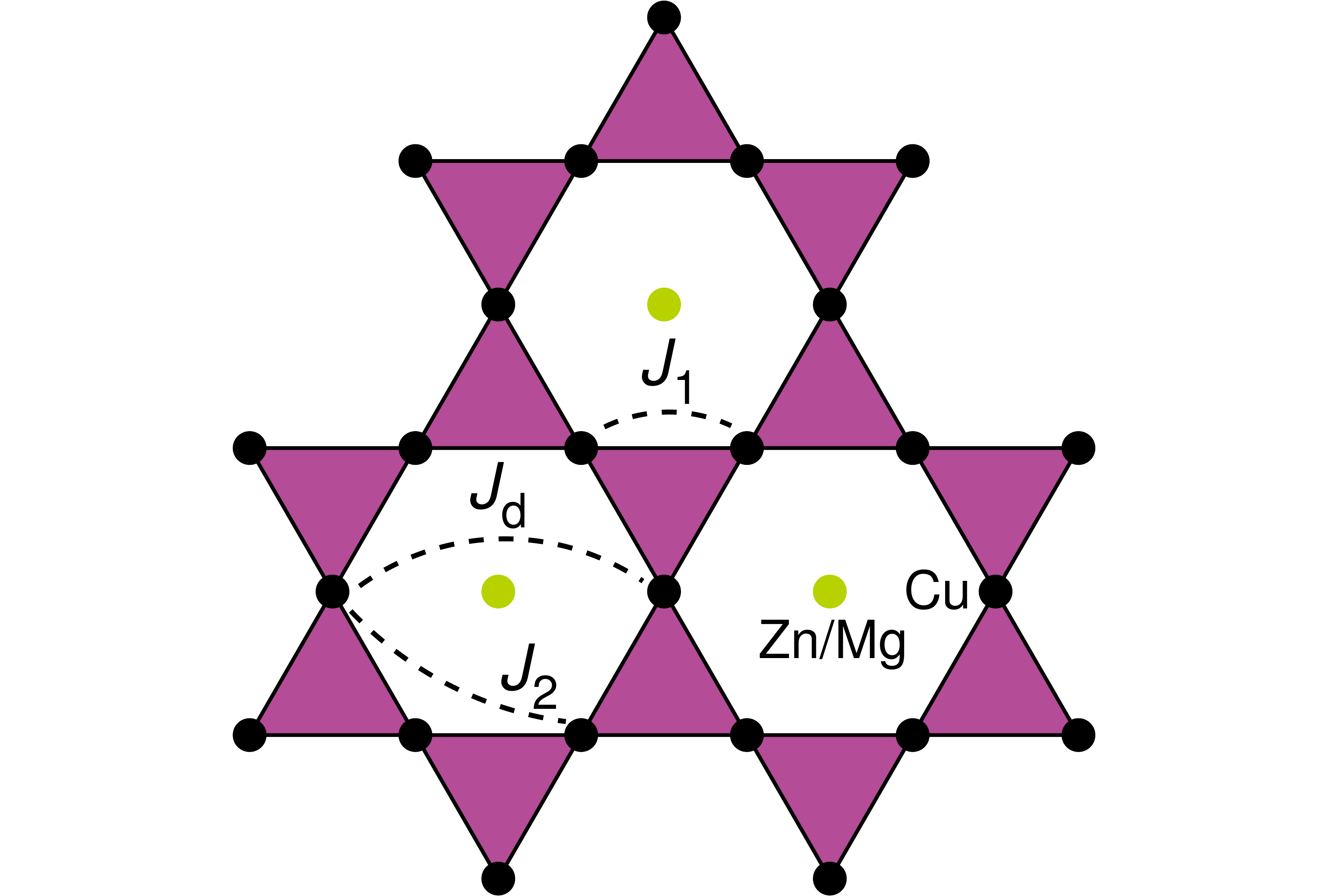}
\caption{For all kagome magnets considered, the lattice is formed by Cu$^{2+}
$ $S$=$\frac{1}{2}$ spins (black), and the Heisenberg exchange couplings are
given by nearest neighbor $J_{1}$, second nearest neighbor $J_{2}$, and
diagonal $J_{d}$ across the hexagons. Kapellasite and haydeeite feature
in-plane Zn$^{2+}$ and $\mathrm{Mg}^{2+}$ ions (green), respectively, at the
center of the hexagons, triggering appreciable values for $J_{d}$.}
\label{fig:lattice}
\end{figure}

{\it Introduction}. Quantum magnetism in low-dimensional systems with parametric or geometrical
frustration has been a highly inspiring field of research ever since the
seminal paper of Pomeranchuk~\cite{Pomeranchuk-1941}. A Holy Grail of the field has been 
the spin-$\frac{1}{2}$ antiferromagnet on the kagome lattice (Fig.~\ref
{fig:lattice}), where the geometrical frustration inherent in the individual
triangles is only marginally alleviated through a corner sharing lattice
pattern~\cite{Balents-2010}. It is widely conjectured to host a spin liquid
phase, albeit the nature and the topological classification of this phase
are still controversial~\cite
{Singh-2007,Ran-2007,Nakano-2011,Lauchli-2011,Iqbal-2011a,*Iqbal-2011b,Yan-2011,Depenbrock-2012,Jiang-2012, Capponi-2013,Iqbal-2013,Xie-2014,Punk-2014,Iqbal-2014}. It may host exotic superconducting phases~\cite{Nandkishore-2013,
Mazin-2014}.

\begin{figure*}
\includegraphics[width=1.0\textwidth]{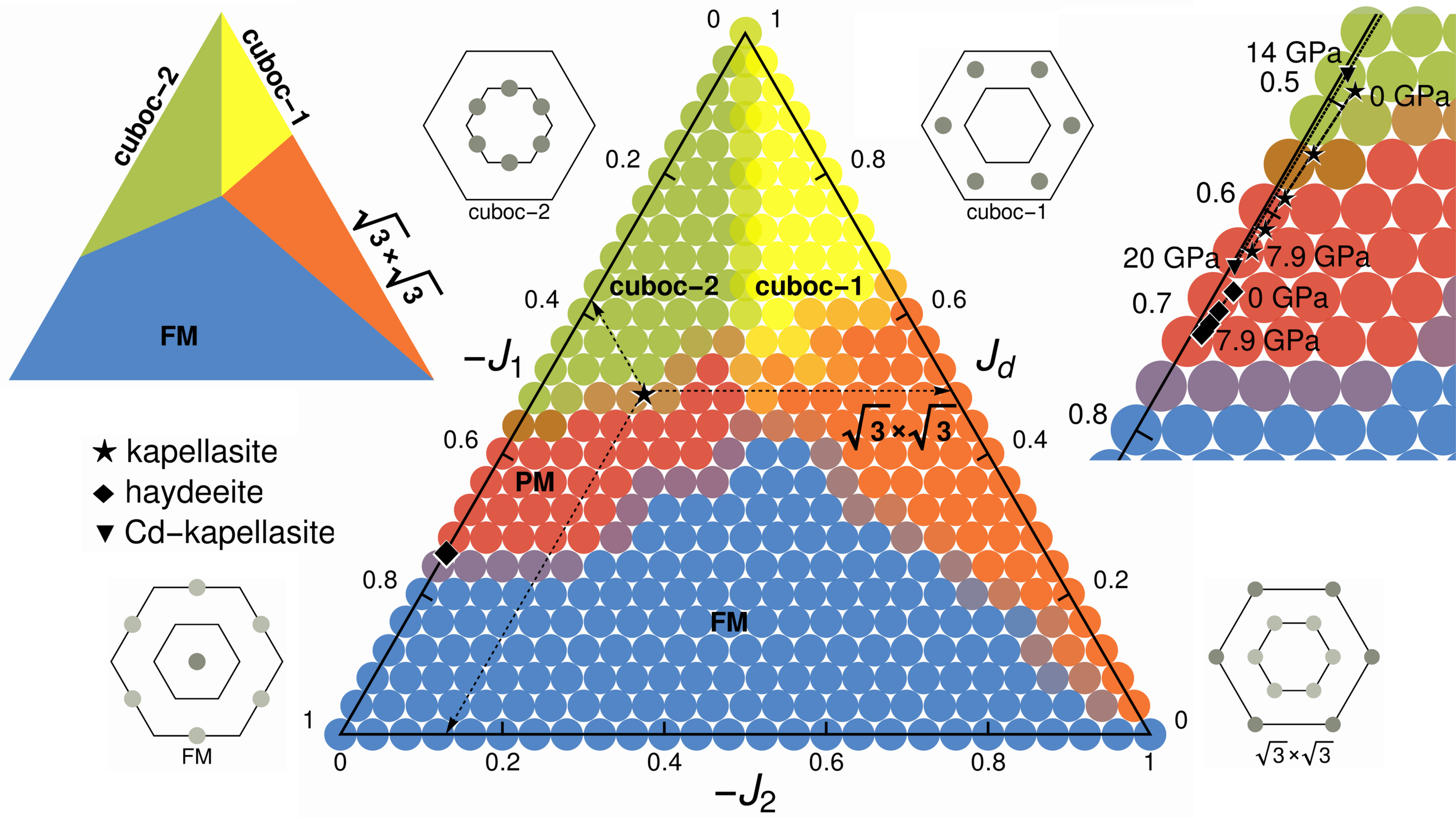}
\caption{Quantum phase diagram of the $J_{1}$-$J_{2}$-$J_{d}$ Heisenberg
model as defined in Eq.~(\ref{eqn:heis-ham}). It features a large
paramagnetic (PM) domain for intermediate $J_{d}$. The exchange couplings
estimated from fitting experimental data for kapellasite and haydeeite are
marked~\protect\cite{Bernu-2013,Boldrin-2015}. The static spin structure
factors in the extended Brillouin zone are shown next to each phase. The
corresponding classical phase diagram is shown in the upper left. The
evolution of the couplings in the different materials under application of
pressure, as calculated by \textit{ab initio} methods, is shown in the
enlarged region on the right. }
\label{fig:phase_diagram}
\end{figure*}

Recently, significant progress has been achieved regarding experimental
realizations of kagome magnets. The couplings in the actual materials, however,
differ significantly from idealized models. The most prominent materials are
herbertsmithite [Zn{Cu$_{3}$(OH)$_{6}$Cl$_{2}$}]~\cite{Shores-2005,Han-2012a}
and its polymorphs, kapellasite [{ZnCu$_{3}$ (OH)$_{6}$Cl$_{2}$}]~\cite
{Krause-2006, Colman-2008} and haydeeite [{MgCu$_{3}$(OH)$_{6}$Cl$_{2}$}]
~\cite{Malcherek-2007,Schlueter-2007,Colman-2010,Chu-2011,Boldrin-2015}, with
ground states ranging from potentially paramagnetic (PM) to weakly
ferromagnetic (FM) phases. This variety of ground states in structurally
similar systems calls for a thorough theoretical investigation of the
experimentally relevant couplings and their implications. In this Rapid Communication,
we focus on the following aspects: (1) Which phases\textemdash ordered or PM\textemdash 
are realized, depending on Heisenberg couplings $J_{1}$, $J_{2}$, and $J_{d}$
(Fig.~\ref{fig:lattice})? (2) What determines the nature of magnetic
interactions, and why are they so different in these compounds? (3) Can we
deliberately manipulate these materials to probe different parts of the
phase diagram, and, in particular, the PM (possibly spin-liquid) phase
indicated in Fig.~\ref{fig:phase_diagram}?

To begin with, we map out the zero-temperature phase diagram of the $J_{1}$-$
J_{2}$-$J_{d}$ kagome model. Specifically, we employ the pseudofermion
functional renormalization group (PF-FRG)~\cite
{Reuther-2010,Reuther-2011a,Suttner-2014} method to compute magnetic
fluctuations. We find that the experimentally estimated parameters place
kapellasite near the borderline between the PM and the antiferromagnetically
ordered cuboc-2 phase and haydeeite near the PM/FM border. Using \textit{ab
initio} density functional theory (DFT) calculations, we then discuss the reliability of these parameters 
and possible microscopic origins for their variations. We proceed with
suggestions on how one can modify these compounds to shift them away from
their current positions and explore other parts of the phase diagram. Among
other aspects, our results provide an independent assessment of the initial
placement of the materials in the phase diagram.

%%%%%%%%%%%%%%%%%%%%%%%%%%%%%%%%%%%%%%%%
%%%%%%%%%%%%%%%%%%%%%%%%%%%%%%%%%%%%%%%%

Herbertsmithite, kapellasite (KL), and haydeeite (HD) feature geometrically
perfect $\mathrm{Cu^{2+}}$ $S=\frac{1}{2}$ kagome planes with the
nearest-neighbor (NN) superexchange $J_{1}$ mediated by OH$^{-}$ and $
\mathrm{Cl^{-}}$. The minimal model also includes subdominant interactions $
J_{2}$ and  $J_{3}$ and significant $J_{d}$ (Fig.~\ref{fig:lattice}). Experiment~\cite
{Han-2012b} and calculations~\cite{Misguich-2007,Valenti-2013} suggest that
in herbertsmithite, where only Cu is present in the kagome planes, both $
J_{3}$ and $J_{d}$ are negligible, and thus the material is well described
by a NN antiferromagnetic model with $J_{1}\approx 180$ K, with a small
but non-negligible $J_{2}$~\cite{Suttner-2014,Iqbal-2015}. The quantum paramagnetic
ground state of such an antiferromagnet has been intensively studied
theoretically (see, e.g., Refs.~\cite
{Singh-2007,Ran-2007,Nakano-2011,Lauchli-2011,Iqbal-2011b,Yan-2011,Depenbrock-2012,Capponi-2013,Iqbal-2012,Iqbal-2013,Xie-2014,Punk-2014,Iqbal-2014,*Becca-2015}), and there is experimental evidence of a spin liquid state in
herbertsmithite~\cite{Mendels-2007,Han-2012a,Fu-2015}.

On the other hand, in KL and HD, the Zn and Mg ions, respectively, occupy the centers
of the Cu hexagons~\cite{Colman-2008}, thus spanning the Cu pairs connected
by $J_{d}$. This seems to explain why the $J_{d}$ interaction is sizable,
and differs considerably between the two compounds. However, the nature of
this interaction is probably more complex than that, as discussed in detail
in the Supplemental Material~\cite{Supp}. Bernu \textit{et al.}~\cite
{Bernu-2013} extracted $J$'s in KL from the temperature dependencies of
magnetic susceptibility and specific heat, while Boldrin \textit{et al.}~
\cite{Boldrin-2015} did the same for HD, using spin-wave dispersion. With
the caveat that these are distinctly different experimental procedures, the
estimated exchange coupling constants are $(J_{1},J_{2},J_{d})=(-12,-4,15.6)$
K for KL~\cite{Bernu-2013} and $(J_{1},J_{d})=(-38,11)$ K, with ${J_{2}}/{
J_{1}}\ll 0.1$ for HD~\cite{Boldrin-2015}. The small and negative values of $
J_{1}$ signal a large cancellation of the anti- and ferromagnetic
contributions to the NN superexchange, which, as we discuss in the
Supplemental Material~\cite{Supp}, is quite unexpected, but that they are
close in both compounds is consistent with their similar geometries. Further
investigations using electron spin resonance estimated the symmetric
exchange anisotropy $D$ to be only $|D/J_{1}|\sim 3\%$~\cite{Kermarrec-2014}
, thus justifying the use of the $(J_{1},J_{2},J_{d})$ \textit{isotropic}
Heisenberg model as a good starting point for both KL and HD.

{\it Model and Methods}. The Heisenberg Hamiltonian reads 
\begin{equation}
\hat{\mathcal{H}}=J_{1}\sum_{\langle ij\rangle }\hat{\mathbf{S}}_{i}\cdot 
\hat{\mathbf{S}}_{j}+J_{2}\sum_{\langle \!\langle ij\rangle \!\rangle }\hat{
\mathbf{S}}_{i}\cdot \hat{\mathbf{S}}_{j}+J_{d}\sum_{\langle \!\langle
\!\langle ij\rangle \!\rangle \!\rangle _{d}}\hat{\mathbf{S}}_{i}\cdot \hat{
\mathbf{S}}_{j},  \label{eqn:heis-ham}
\end{equation}%
where $\hat{\mathbf{S}}_{i}$ is the spin-$\frac{1}{2}$ operator at site $i$.
Here, $J_{1}$, $J_{2}\leqslant 0$ (ferromagnetic) and $J_{d}\geqslant 0$
(antiferromagnetic), normalized so that $|J_{1}|+|J_{2}|+J_{d}=1$. The
indices $\langle ij\rangle $, $\langle \!\langle ij\rangle \!\rangle $, and $
\langle \!\langle \!\langle ij\rangle \!\rangle \!\rangle _{d}$ denote sums
over NN and next-NN bonds, and the diagonals of hexagons, respectively
(Fig.~\ref{fig:lattice}).

% %%%%%%%%%%%%%%%%%%%%%%%%%%%%%%%%%%%%%%%%%%%%%
%%%%%%%%%%%%%%%%%%%%%%%%%%%%%%%%%%%%%%%%

\begin{figure}
\includegraphics[width=1.0\columnwidth]{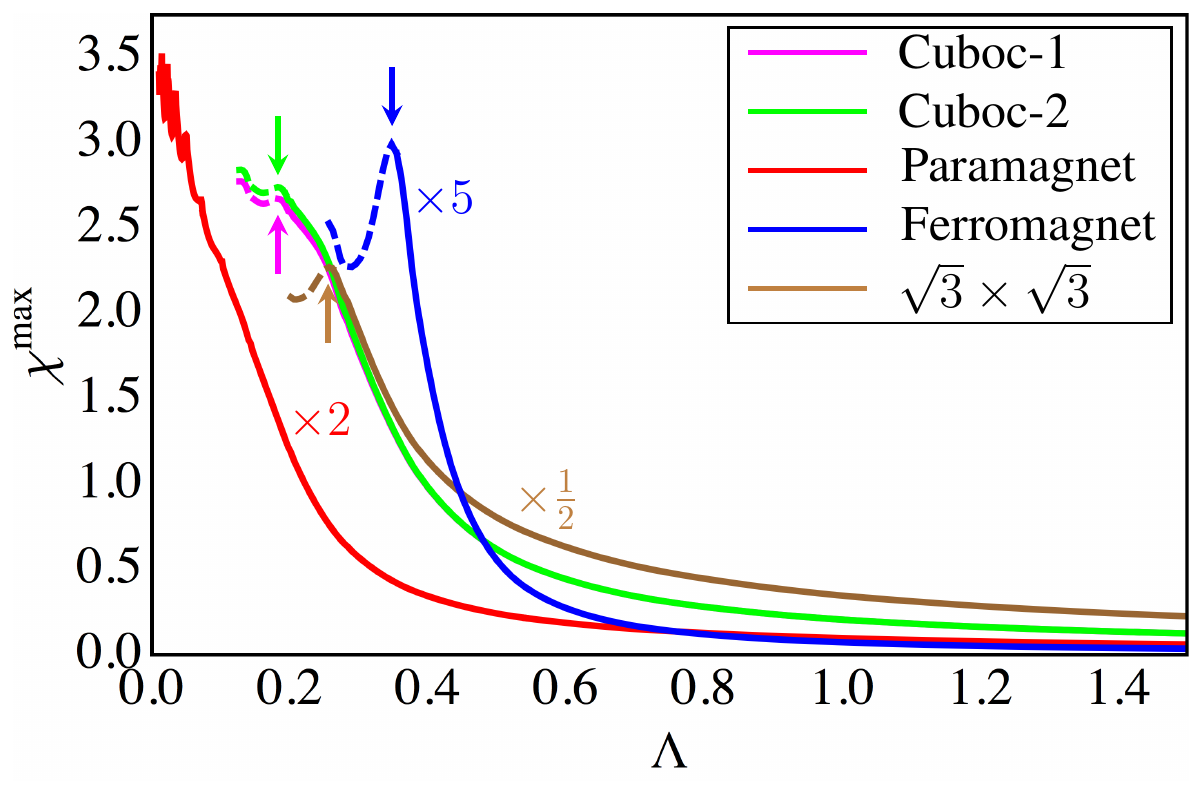}
\caption{Representative RG flows of the magnetic susceptibilities at the
ordering wave vectors of the four ordered regimes of Fig.~\protect\ref
{fig:phase_diagram} and the paramagnetic regime. The points at which the
solid lines become dashed (marked by arrows) indicate an instability in the
flow and express the onset of order. In the smooth flow (red curve)
indicating paramagnetism, no such instability is found. The small
oscillations below $\Lambda\approx0.1$ in this flow are due to frequency
discretization.}
\label{fig:rgflow}
\end{figure}

\begin{figure*}
\includegraphics[width=1.0\textwidth]{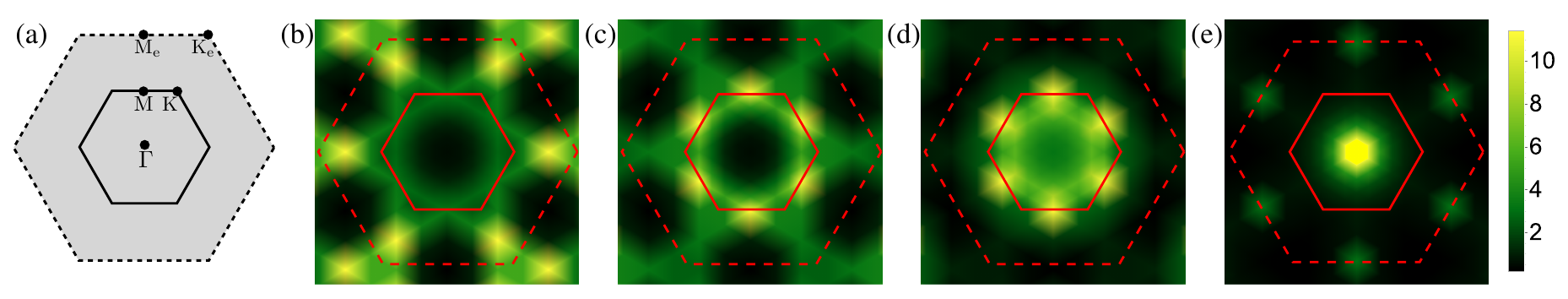}
\caption{ (a) The first (solid line) and extended (dashed line) Brillouin
zones of the kagome lattice. (b)--(e) Representative spin susceptibility
profiles obtained in PF-FRG for different regimes of the quantum phase
diagram in Fig.~\protect\ref{fig:phase_diagram}: (b) cuboc-1, (c) cuboc-2,
(d) paramagnetic (PM), (e) ferromagnetic (FM).}
\label{fig:sq}
\end{figure*}

In the PF-FRG approach~\cite
{Reuther-2010,Reuther-2011a,Reuther-2011b,Reuther-2011c,Suttner-2014}, we
first rewrite Eq.~(\ref{eqn:heis-ham}) in terms of pseudofermions as $\hat{
\mathbf{S}}_{i}=\frac{1}{2}\sum_{\alpha \beta }\hat{c}_{i,\alpha }^{\dagger }
\boldsymbol{\sigma }_{\alpha \beta }\hat{c}_{i,\beta }$, ($\alpha ,\beta
=\uparrow ,\downarrow $), where $\hat{c}_{i,\alpha }$ are the pseudofermion
operators, and $\boldsymbol{\sigma }$ are Pauli matrices. This enables us to
apply Wick's theorem and develop a diagrammatic technique. To this end, an
infrared frequency cutoff $\Lambda $ is introduced in the fermion
propagator. The FRG ansatz (for recent reviews, see, e.g., Refs.~\cite
{Metzner-2012,Platt-2013}) then formulates an infinite hierarchy of coupled
integrodifferential equations for the evolution of all $m$-particle vertex
functions under the flow of $\Lambda $. Within PF-FRG, the truncation of
this system of equations to a closed set is accomplished by the inclusion of
only two-particle reducible two-loop contributions, which ensures sufficient
backfeeding of the self-energy corrections to the two-particle vertex
evolution~\cite{Katanin-2004}. A crucial advantage of the PF-FRG is that the
diagrammatic summation incorporates vertex corrections between all
interaction channels, i.e., treats magnetic ordering and disordering
tendencies on equal footing. The PF-FRG equations are solved numerically by
discretizing the frequency dependencies of the vertex functions and limiting
the spatial dependencies to a finite cluster. We used 64 discretized
frequencies and a cluster of $432$ sites. In the PF-FRG approach, the onset of magnetic long-range order is
signaled by a breakdown of the smooth RG flow, whereas a smooth evolution
down to $\Lambda \rightarrow 0$ (where $\Lambda $ is the infrared frequency
cutoff) indicates PM behavior~\cite{Reuther-2010} (Fig.~\ref
{fig:rgflow}). From the effective low-energy two-particle vertex, we obtain
the spin-spin correlation function in real space, which we then convert into
the momentum-resolved spin susceptibility (see Fig.~\ref{fig:sq}).

Previous applications of the PF-FRG method to frustrated magnets have been
extremely successful. In particular, (i) the determined magnetic and
nonmagnetic phases of the spin-$\frac{1}{2}$ Heisenberg $J_{1}$-$J_{2}$
antiferromagnet on the square lattice and the locations of the phase
transitions  quantitatively agree with DMRG, exact diagonalization, and
other methods~\cite{Reuther-2010}, (ii) the phase diagram of the $J_{1}$-$
J_{2}$-$J_{3}$ Heisenberg model on the honeycomb lattice agrees perfectly
with exact diagonalization~\cite{Reuther-2011b,Albuquerque-2011}, (iii) the
phase diagram of the Kitaev-Heisenberg model is also correctly determined within the
PF-FRG~\cite{Reuther-2011c,Chaloupka-2010}, in particular, the short range
nature of the spin correlations in the Kitaev limit is correctly reproduced, and
(iv) the spin structure factor of the NN Heisenberg antiferromagnet on the
Kagome lattice in PF-FRG is in very good quantitative agreement with DMRG~
\cite{Suttner-2014,Depenbrock-2012}, which is of particular relevance to the
problem at hand.

For \textit{ab initio} DFT calculations we used the generalized gradient
approximation (GGA) functional~\cite{GGA}. Structure optimizations were
performed with the projector augmented wave method within the VASP code~\cite{Kresse-1993,Kresse-1996}, and accurate total energies were calculated using the all-electron FPLO code~\cite{Koepernik-1999,*FPLO-webpage}.

%%%%%%%%%%%%%%%%%%%%%%%%%%%%%%%%%%%%%%%%
%%%%%%%%%%%%%%%%%%%%%%%%%%%%%%%%%%%%%%%

{\it Results}. 
The PF-FRG quantum phase diagram of the $J_{1}$-$J_{2}$-$J_{d}$ model of Eq.~
(\ref{eqn:heis-ham}) is depicted in Fig.~\ref{fig:phase_diagram}. Individual
data points are labeled according to which type of phase they belong to in the 
PF-FRG. For small $J_{d}$, FM dominates. For intermediate $J_{d}$ and large $
J_{2}$, a $\sqrt{3}\times \sqrt{3}$ order is found which changes into the
cuboc-1 order for increasing $J_{d}$. The cuboc orders describe
12-sublattice noncoplanar orders in which the spins orient towards the
corners of a cuboctahedron~\cite{Messio-2011,Gong-2015} (that is, along the
12 possible [110] directions). For the domains discussed so far, the quantum
phase diagram approximately matches the classical phase diagram of Eq.~(\ref
{eqn:heis-ham}) (Fig.~\ref{fig:phase_diagram}). Quantum corrections start
to become visible closer to the cuboc-1/cuboc-2 boundary, as ${J_{2}}$
becomes smaller than ${J_{1}}$ (for large $J_{d})$. The classical first
order transition line between the cuboc-1 and cuboc-2 phases is then
replaced by a narrow vertical strip ($J_{1}\approx J_{2}$) in the quantum
case, depicted by a merging gradient in Fig.~\ref{fig:phase_diagram}, where
an effectively $1D$ paramagnetic chain regime is found. As the most
important modification to the classical picture, however, an extended PM
regime emerges for ${J_{2}}/{J_{1}}<1$ separating the cuboc-2 from the FM
domain. Its spin susceptibility profile has a well-defined wave-vector
dependence featuring dominant short-distance correlations with soft maxima
at cuboc-2 ordering wave vectors and subdominant FM correlations. This
type of magnetic fluctuation profile is rather peculiar for a PM phase and
fundamentally different from what is found for herbertsmithite~\cite
{Han-2012b,Suttner-2014}. As we enter the PM regime from the cuboc-2 phase
by lowering $J_{d}$, the magnetic correlations change quantitatively but not
qualitatively, as is manifest from a comparison of their spin susceptibility
profiles in Fig.~\ref{fig:sq}(c) and Fig.~\ref{fig:sq}(d). The only notable
difference is more spectral weight smearing in the PM regime. Within the PM
regime, the spin correlations are found to be short-ranged, and calculations
of the dimer response function rule out any kind of valence-bond crystal
order up to a $36$-site cell. Note that the $J_{1}$-$J_{2}$-$J_{d}$ model
has been recently analyzed by variational Monte Carlo (VMC) \cite{Bieri-2015}
. There, in the large $J_{d}$ regime, noncoplanar cuboc-1 order is absent,
and cuboc-2 order is reduced to a small part of the parameter space.
Instead, for a significant range of $J_{d}$ and depending on  ${J_{2}}/{J_{1}
}$, two distinct gapless $U(1)$ chiral spin liquids with a spinon Fermi
surface are found over an extended region. As opposed to the PF-FRG, which
treats magnetic order and disorder tendencies on the same footing, 
a certain bias of VMC against cuboc-1 and cuboc-2 orders can be argued for
on the basis of the variational wave functions employed. The noncoplanar
structure of cuboc-1 and cuboc-2 orders implies that the corresponding
chosen Jastrow wave functions are inaccurate as the Jastrow factor does not
correctly describe the relevant quantum fluctuations on top of the classical
state~\cite{Bieri-2015}. On the other hand, PF-FRG does not suffer from this
deficiency, and, if anything, may slightly \textit{over}estimate the PM
domain.

The location of the high-temperature series expansion estimate of exchange
couplings for KL~\cite{Bernu-2013} is marked within the center triangle by a
star at $(0.38,0.13,0.49)$ in Fig.~\ref{fig:phase_diagram}. This is very
close to the boundary between the cuboc-2 and the quantum PM phases.
Experimentally, KL shows no spin freezing and persistent fluctuations down
to 20 mK (by $\mu {\rm SR})$, a diffused continuum of excitations (inelastic
neutron scattering), and the divergence of the intrinsic local
susceptibility for $T\rightarrow 0$ in NMR~\cite{Fak-2012}. The static spin
structure factor shows a well-defined wave-vector dependence exhibiting AFM
short-range correlations~\cite{Janson-2008,Messio-2011}, consistent with the
cuboc-2 pattern. This whole set of experiments has been interpreted in favor
of a gapless quantum spin liquid ~\cite{Fak-2012} close to the cuboc-2 AFM
order. The delicate location of KL might have important experimental
implications, in that only moderate modifications in material synthesis or
experimental conditions amounting to strain, pressure, defects/impurities,
and the imminent presence of different type of anisotropies would lead to
significant effects. Slight modifications of the Heisenberg coupling
constants could drive kapellasite either into a weak cuboc-2 order or
towards a quantum PM phase. As tentatively observed in current experiments,
this finding is consistent with the compound exhibiting strong magnetic
frustration and significant ordering fluctuation tendencies towards cuboc-2
at the same time. Our PF-FRG calculations show that this kapellasite
location yields a critical flow, that is, neither shows a robust and smooth
RG flow down to $\Lambda =0$ that would point at quantum PM nor exhibits a
clear signature of an order-induced breakdown.

The location of the linear spin-wave estimate of exchange couplings for
haydeeite~\cite{Boldrin-2015} is marked by the diamond at $(0.77,0.0,0.23)$
in Fig.~\ref{fig:phase_diagram}. Remarkably, it is likewise located at the
border with the PM regime, but now on the FM side. Experiments~\cite{Boldrin-2015}
suggest a very weak FM order below $4.2$ K. From our PF-FRG analysis,
ferromagnetic order is weak but unambiguous, and the magnetic fluctuations
clearly show FM signatures: a dominant peak at $\Gamma $ and subdominant
peaks at $\mathrm{M_{e}}$ [see Fig.~\ref{fig:sq}(e)].

To summarize, the reported exchange parameters~\cite{Bernu-2013,Boldrin-2015}
place KL and HD on opposite sides of the (arguably most interesting)
paramagnetic region. Moreover, it seems like these two compounds,
accidentally, are both at or very close to a borderline between two phases.
While making them especially intriguing, it also considerably complicates
their study. It is therefore highly desirable to be able to modify the same
compounds continuously, in order to \textquotedblleft
traverse\textquotedblright\ the phase diagram. In principle, there are
several ways to do so.

One option is alloying Mg and Zn by creating a mixed compound Mg$_{x}$Zn$
_{1-x}${Cu$_{3}$(OH)$_{6}$Cl$_{2}$} as suggested in Ref.~\cite{Boldrin-2015}
. However, while this proposal creates a system with \textit{average}
exchange couplings intermediate between those in KL and HD, in reality it
will have random bonds with exchange constants similar to those either in KL
or HD and is more likely to freeze into a spin glass state rather than to
develop a spin-liquid phase. Besides, chemical substitution of Zn by Mg
naturally affects $J_{d}$, but the effect on $J_{1}$ is harder to predict.

%%%%%%%%%%%%%%%%%%%%%%%%%%%%%%%%%%%%%%%%%%%%%
%%%%%%%%%%%%%%%%%%%%%%%%%%%%%%%%%%%%%%%%%%%%%

These considerations lead us to propose alternative options without
introducing additional disorder. For that, we need first to exercise some
caution when using the experimental numbers. Indeed, Refs.~\cite
{Bernu-2013,Boldrin-2015} are complementary in terms of methodology used to
extract the exchange coupling constants; Ref.~\cite{Bernu-2013} relies on
magnetometry/calorimetry while Ref.~\cite{Boldrin-2015} does a spin-wave
analysis. In both cases, and especially in HD, actual samples have
considerable excess of Zn or Mg, substituting for Cu. Each missing Cu
creates four incompletely frustrated spins, which may alter the results
compared to the stoichiometric compound. While the reported parameters for
HD are consistent with a FM ground state, the experimental data~\cite
{Boldrin-2015} looks more complicated than that. Indeed, the ordered moment
from neutron scattering was estimated to be $\lesssim 0.2\mu _{\mathrm{B}}$~
\cite{Boldrin-2015}, while the saturation moment from magnetization was $0.83
$~\cite{Boldrin-2015} or $1.0\mu _{\mathrm{B}}$~\cite{Chu-2011}, and the
spontaneous ordered moment is $0.02\mu _{\mathrm{B}}$~\cite{Colman-2010} or
less~\cite{Chu-2011}. The Curie-Weiss effective moment is $1.83\mu _{\mathrm{
B}}$~\cite{Colman-2010}, consistent with an ordered moment greater than $
1\mu _{\mathrm{B}}$. A 40 times difference between the saturation and
spontaneous magnetization is highly unusual, and so is the discrepancy
between spontaneous and neutron-measured moments. While this has been
vaguely ascribed to frustration~\cite{Boldrin-2015}, the \textquotedblleft
frustration parameter\textquotedblright , usually defined as the ratio of
the mean field $T_{\mathrm{CW}}$ and frustration-suppressed $T_{\mathrm{N}}$
is in fact less than one here. The magnetic susceptibility starts growing
with cooling below 5 K, but does not diverge at the putative $T_{\mathrm{C}
}=4.2$ K, and instead starts flattening out below 4 K. Overall, the magnetic
behavior described in Ref.~\cite{Boldrin-2015} is more typical for canted
antiferromagnets than for ferromagnets. On the other hand, an independent
study~\cite{Chu-2011} found much larger magnetic moment and stronger
ferromagnetic behavior.

Given the experimental situation, we decided to address the question
theoretically by performing DFT+$U$ calculations (see Supplemental Material). We
used the FPLO program with $U=6-8$ eV, using the fully-localized double
counting scheme, orthogonal projection of $3d$ density, and gradient
corrections in the DFT functional, and we found that, in agreement with 
previous observations~\cite{Valenti-2013}, this setup gives the closest
agreement with the experiment for KL. As discussed in detail in the
Supplemental Material, the exchange coupling constants in these materials
are not only small, but they also depend on the technical details of the setup
(we remind the reader that while DFT is a first principles method, DFT+$U$ is
not), which accounts, for instance, for the difference between Refs.~\cite
{Valenti-2013} and~\cite{Janson-2008}. On the other hand, the trends in the
dependence of the exchange parameters on the geometry and chemistry in these
compounds are quite robust, and therefore such calculations can be used as a
guidance for modifying existing materials in order to steer them toward one
or another magnetic phase.

%%%%%%%%%%%%%%%%%%%%%%%%%%%%%%%%%%%%%%%%%%%%%
\begin{figure}[b]
\includegraphics[width=1.0\columnwidth]{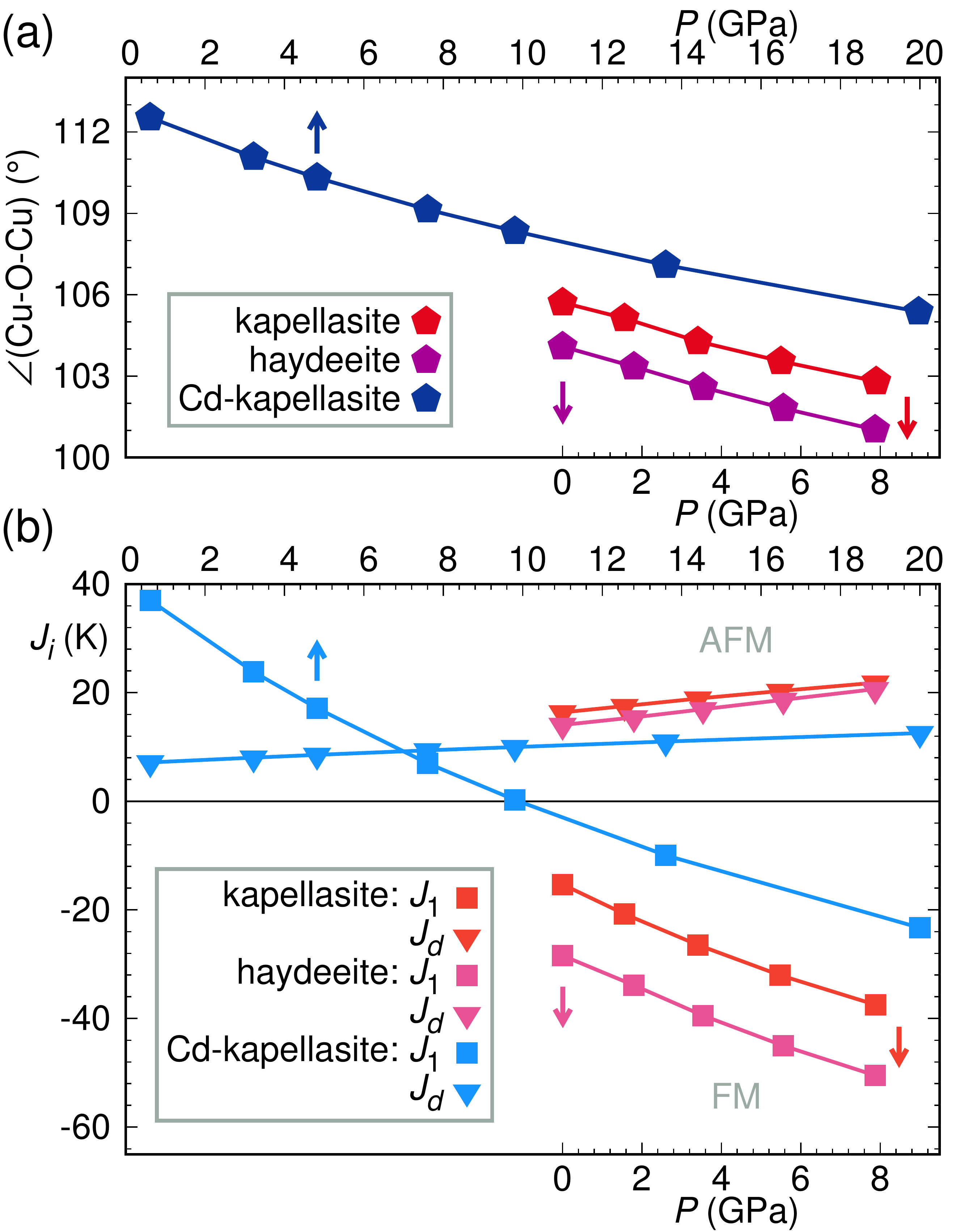}
\caption{ (a) Predicted evolution of Cu-O-Cu angles as a function of pressure
for kapellasite, haydeeite, and the hypothetical Cd-kapellasite. (b) Pressure
induced changes in the two dominant exchange coupling parameters $J_1$ and $
J_d$ for the three compounds.}
\label{fig:pressure}
\end{figure}
%%%%%%%%%%%%%%%%%%%%%%%%%%%%%%%%%%%%%%%%%%%%%

With this setup, using reported crystal structures (and optimized positions
of hydrogen, since these are not known experimentally) we obtained for KL ($
J_{1},J_{2},J_{d})=(-12.5,-0.55,16.1)$K, corresponding to $(0.43,0.02,0.55)$
. This is rather close to $(-12,-4,15.6)$K~\cite{Bernu-2013}, albeit a bit
deeper in the cuboc-2 phase. For HD we find $(-21.2,0.57,12.8)$K, or $
(0.61,0.02,0.37)$, placing it in the PM regime (see inset in Fig.~\ref
{fig:phase_diagram}). By comparing calculations for the same structure, but
substituting Zn for Mg, or for the same composition but different structure,
we found, not surprisingly, that $J_{1}$ is predominantly (80--90\%)
controlled by the structure and $J_{d}$ by the bridging element. The
smaller Cu-O-Cu angle in HD of 104.98$\degree$ vs 105.84$\degree$ in KL
results in a larger value for $J_{1}$, while the additional hopping path 
\emph{via} semicore Zn $3d$ states provides a larger $J_{d}$. The principal
discrepancy with the experimental values appears to be the overestimation of 
$J_{d}$ in HD. Regardless of whether this is an experimental problem (\textit{e.g.}, imperfect samples) or theoretical (\textit{e.g.},
overestimation of Mg-O $sp\sigma $ hopping), the trends in the dependence of
the exchange parameters with respect to both structural and chemical changes
are well reproduced. Having identified the origin of the behavior of $J_{1}$
and $J_{d}$, we propose two recipes for sampling the phase diagram.

The first option is to apply pressure, keeping the chemical compositions. To
investigate this avenue, we have calculated the structures of KL and HD at
experimental volumes and at compressions up to 12{\%} (Fig.~\ref
{fig:pressure}). These compressions correspond to pressures of $\approx 7.9$~GPa for both KL and HD, which is experimentally accessible. In both cases,
the Cu-O-Cu angle decreases systematically, and $J_{1}$ increases by up to
140{\%} (KL) and 80{\%} (HD), as shown in Fig.~\ref{fig:pressure}. Thus, by
applying pressure we should be able to drive KL into the PM regime, and even
approach the boundary with the FM phase, while applying pressure to HD will
drive it deeper into the FM regime.

The second option combines both structural and chemical changes. One can
preserve homogeneity by substituting Zn in KL with Cd. Due to the larger
ionic radius, this substitution may be difficult to realize and might
require high-pressure synthesis. Indeed, our calculations show that while
such a compound would be locally stable, the Cu-Cd plane would be
considerably expanded and the Cu-OH-Cu angle would flatten to $112\degree$--$113\degree$, which renders $J_{1}$ antiferromagnetic~\cite{Supp}. This compound, however, would also be highly compressible and
would return to a structure similar to KL at a pressure of about 20 GPa. One
expects that $J_1$ should already be ferromagnetic at this pressure, while
at the same time, we note that since the $5d$ level in Cd lies considerably
lower compared to the $4d$ level in Zn, the $J_d$ is expected to reduce. Both
conjectures are confirmed by our calculations. Indeed, at $P=20$ GPa the
representative point in the phase diagram appears close to HD at $P=0$,
while at $P=13.6$ GPa it is close to KL at $P=0$ (Fig.~\ref
{fig:phase_diagram}). Thus, synthesizing CdCu$_{3}$(OH)$_6$Cl$_2$ and
applying external pressure provides us a vehicle to traverse a vast extent
of the phase diagram encompassing a large range of $J_{d}$, especially deep
into the paramagnetic phase.

{\it Acknowledgments}. We thank F. Becca, S. Bieri, V. Khanna, C. Lhuillier, T. Neupert, D.
Poilblanc, H. Rosner, and R. Suttner for useful discussions. The work was
supported by the European Research Council through
ERC-StG-TOPOLECTRICS-Thomale-336012. I.I.M. was also supported by the Office of
Naval Research through the Naval Research Laboratory's Basic Research
Program. H.O.J. and R.V. thank the DFG (Deutsche Forschungsgemeinschaft) for
financial support through SFB/TRR 49 and FOR 1346, and Y.I., M.G., and R.T. received
financial support through SFB 1170. J.R. is supported by the Freie Universit\"at Berlin within the Excellence
Initiative of the German Research Foundation.



\section*{Supplementary material}

# Paramagnetism in the kagome compounds $(Zn,Mg,Cd)Cu_3(OH)_6Cl_2$

Yasir Iqbal,[1] Harald O. Jeschke,[2] Johannes Reuther,[3,4] Roser Valentí,[2] I. I. Mazin,[5] Martin Greiter,[1] and Ronny Thomale[1]

[1]*Institute for Theoretical Physics and Astrophysics, Julius-Maximilian's University of Würzburg, Am Hubland, D-97074 Würzburg, Germany*
[2]*Institut für Theoretische Physik, Goethe-Universität Frankfurt, Max-von-Laue-Straße 1, D-60438 Frankfurt am Main, Germany*
[3]*Dahlem Center for Complex Quantum Systems and Fachbereich Physik, Freie Universität Berlin, D-14195 Berlin, Germany*
[4]*Helmholtz-Zentrum Berlin für Materialien und Energie, D-14109 Berlin, Germany*
[5]*Center for Computational Materials Science, Naval Research Laboratory, Code 6390, 4555 Overlook Ave, SW, Washington, DC 20375, United States of America*

## I. DENSITY FUNCTIONAL THEORY CALCULATIONS

### A. Prediction of high pressure structures

We predict high pressure structures using the Vienna ab initio simulation package (VASP) implementation [1, 2] of DFT within the generalized gradient approximation (GGA) [3]. We base our calculations on the structures of Krause *et al.* [4] for kapellasite, and of Malcherek *et al.* [5] for haydeeite. We obtain a $P = 0$ GPa relaxed structure by fixing the experimental volume and relaxing all lattice and internal structural parameters. The DFT calculated pressure at this volume, due to the well known underbinding of the GGA exchange correlation functional, is $P_{\text{offset}} = 1.54$ GPa for kapellasite and $P_{\text{offset}} = 0.93$ GPa for haydeeite. We then proceed to reduce the volume in steps of 3%, relaxing all lattice and internal structural parameters. The physical pressure values are obtained by subtracting $P_{\text{offset}}$ from the calculated pressure at each reduced volume.

### B. Determination of Hamiltonian parameters

The calculations were performed with the full potential local orbital (FPLO) basis set [6] and the GGA functional [3]. Total energies for sets of different spin configurations were calculated with GGA+U using the atomic limit double counting correction [7], and we employed orthogonal projection of the Cu $3d$ densities. We would like to note that the results are very sensitive to (i) the choice of exchange correlation functional (LDA or GGA), (ii) the double counting correction in the LDA+U approach, and (iii) the choice of projection scheme for the Cu $3d$ orbitals in the FPLO basis. In Ref. [8] the calculations were performed with the LDA exchange correlation functional and the around mean field double counting correction was chosen for the LDA+U functional. With this setup the ferromagnetic contribution in $J_1$ for kapellasite is strongly suppressed, however, with the same setup the results of our two groups agree [9].

We fix $J_{\text{H}} = 1$ eV and vary $U$. Tables I and II show the result of fitting to 9 spin configurations of an orthorhombic $\sqrt{2} \times \sqrt{2} \times 1$ supercell with $P1$ space group, containing 6 inequivalent copper ions. A $6 \times 6 \times 6$ $k$ mesh was used after confirming that an $8 \times 8 \times 8$ $k$ mesh yields identical results. The error given in Tables I and II reflects the statistical error of fitting 9 energies to 5 unknowns (reference energy and four exchange couplings). We have also performed calculations with $2 \times 2 \times 1$ supercells of kapellasite and haydeeite yielding 46 distinct total energies from 12 inequivalent copper ions. The two approaches proved to be precisely equivalent so that the computationally less demanding smaller supercell was chosen for the final computations. Very small fitting errors indicate that the $Cu^{2+}$ spins are well localized and their interaction is very well described by the chosen Heisenberg model.

The two significant exchange couplings, $J_1$ and $J_d$, obtained in this way, are shown visually in Fig. 1 for kapellasite and in Fig. 2 for haydeeite. Evolution of couplings with pressure, and with onsite interaction $U$ is very smooth. The ferromagnetic nearest neighbour coupling in the kagome plane is larger in haydeeite compared to kapellasite while the antiferromagnetic exchange $J_d$ coupling $Cu^{2+}$ ions across the hexagons of the kagome lattice is slightly larger in kapellasite than in haydeeite. $J_1$ varies more strongly with pressure than $J_d$, reflecting its sensitivity to the bond angle. Increasing the onsite interaction $U$ from $U = 6$ eV to $U = 7$ eV and $U = 8$ eV leads to a smooth decrease in absolute value of both exchange couplings. The other two exchange couplings are not shown in the figure; $J_2$ is very small at $\sim 0.5$ K and insensitive to pressure, while $J_4$ is nearly exactly zero within computational error.

TABLE I: **Kapellasite exchange couplings.** All structures except the one marked "exp." are predicted as described in the text.

| $P$ (GPa) | $J_1$ | $J_2$ | $J_4$ | $J_d$ |
|---|---|---|---|---|
| GGA+U, $U = 8$ eV, $J_H = 1$ eV | | | | |
| 0 (exp.) | -12.59(3) | -0.55(3) | -0.17(3) | 16.16(4) |
| 0 | -15.27(2) | -0.55(2) | -0.08(3) | 16.40(4) |
| 1.6 | -20.78(2) | -0.50(2) | -0.04(3) | 17.57(4) |
| 3.4 | -26.41(2) | -0.48(2) | 0.01(3) | 18.87(4) |
| 5.5 | -32.02(2) | -0.49(2) | 0.04(3) | 20.28(4) |
| 7.9 | -37.48(2) | -0.53(2) | 0.07(2) | 21.80(3) |
| GGA+U, $U = 7$ eV, $J_H = 1$ eV | | | | |
| 0 | -16.65(4) | -0.56(4) | -0.02(4) | 19.89(6) |
| 1.6 | -22.94(4) | -0.50(4) | 0.03(4) | 21.33(6) |
| 3.4 | -29.37(3) | -0.48(3) | 0.09(4) | 22.95(5) |
| 5.5 | -35.79(3) | -0.48(3) | 0.13(4) | 24.71(5) |
| 7.9 | -42.06(3) | -0.52(3) | 0.16(4) | 26.58(5) |
| GGA+U, $U = 6$ eV, $J_H = 1$ eV | | | | |
| 0 | -18.05(6) | -0.53(6) | 0.10(7) | 24.00(9) |
| 1.6 | -25.22(6) | -0.46(6) | 0.17(6) | 25.80(9) |
| 3.4 | -32.57(5) | -0.42(5) | 0.23(6) | 27.81(9) |
| 5.5 | -39.93(5) | -0.43(5) | 0.28(6) | 29.98(8) |
| 7.9 | -47.13(6) | -0.46(6) | 0.28(7) | 32.33(10) |

TABLE II: **Haydeeite exchange couplings.** All structures except the one marked "exp." are predicted as described in the text.

| $P$ (GPa) | $J_1$ | $J_2$ | $J_4$ | $J_d$ |
|---|---|---|---|---|
| GGA+U, $U = 8$ eV, $J_H = 1$ eV | | | | |
| 0 (exp.) | -21.27(2) | -0.57(2) | -0.13(3) | 12.72(4) |
| 0 | -28.39(2) | -0.55(2) | -0.02(3) | 14.06(4) |
| 1.8 | -33.94(2) | -0.51(2) | 0.03(2) | 15.42(3) |
| 3.5 | -39.50(2) | -0.50(2) | 0.08(2) | 16.97(3) |
| 5.6 | -45.07(2) | -0.51(2) | 0.12(2) | 18.71(3) |
| 7.9 | -50.50(3) | -0.52(3) | 0.08(3) | 20.65(4) |
| GGA+U, $U = 7$ eV, $J_H = 1$ eV | | | | |
| 0 | -31.23(3) | -0.57(3) | 0.05(4) | 16.60(5) |
| 1.8 | -37.57(3) | -0.53(3) | 0.11(3) | 18.26(5) |
| 3.5 | -43.92(3) | -0.52(3) | 0.16(3) | 20.14(4) |
| 5.6 | -50.30(3) | -0.53(3) | 0.21(3) | 22.25(4) |
| 7.9 | -56.56(2) | -0.57(2) | 0.25(3) | 24.59(4) |
| GGA+U, $U = 6$ eV, $J_H = 1$ eV | | | | |
| 0 | -34.15(5) | -0.57(5) | 0.17(5) | 19.52(7) |
| 1.8 | -41.38(6) | -0.52(6) | 0.25(7) | 21.52(9) |
| 3.5 | -48.64(4) | -0.50(4) | 0.31(5) | 23.81(6) |
| 5.6 | -55.94(4) | -0.52(4) | 0.36(4) | 26.36(6) |
| 7.9 | -63.13(4) | -0.57(4) | 0.40(4) | 29.18(6) |

## II. ORIGIN OF EXCHANGE PARAMETERS

$Cu^{2+}$ is usually a textbook case for the Hubbard model. Cu electrons are strongly localized with hopping $t \ll U$, and the standard superexchange perturbation theory works very well. Most experimental papers interpret haydeeite (HD) and kapellasite (KL) in terms of this theory, specifically, in terms of the Goodenough-Kanamori-Anderson rules, which stipulate that at the bond angle $\phi = 90°$, the superexchange process Cu-O-Cu is completely suppressed and the only remaining interaction is ferromagnetic and generated by the Hund's rule coupling on oxygen (the fact

TABLE III: **Hypothetical Cd-kapellasite exchange couplings.** All structures are predicted as described in the text.

| $P$ (GPa) | $J_1$ | $J_2$ | $J_4$ | $J_d$ |
|---|---|---|---|---|
| | GGA+U, $U = 8$ eV, $J_H = 1$ eV | | | |
| 0.6 | 36.96(5) | -0.10(3) | -0.46(5) | 7.15(7) |
| 3.2 | 23.82(4) | -0.09(2) | -0.33(5) | 8.04(6) |
| 4.8 | 17.13(4) | -0.08(2) | -0.27(4) | 8.55(6) |
| 7.6 | 6.98(3) | -0.08(2) | -0.20(4) | 9.38(5) |
| 9.8 | 0.25(3) | -0.08(2) | -0.15(4) | 9.99(5) |
| 13.6 | -9.95(3) | -0.08(2) | -0.11(3) | 11.00(5) |
| 20 | -23.31(3) | -0.09(2) | -0.11(3) | 12.54(4) |
| | GGA+U, $U = 7$ eV, $J_H = 1$ eV | | | |
| 0.6 | 42.37(7) | -0.10(4) | -0.46(8) | 8.77(11) |
| 3.2 | 27.39(6) | -0.08(3) | -0.31(7) | 9.90(9) |
| 4.8 | 19.74(5) | -0.08(3) | -0.25(6) | 10.53(9) |
| 7.6 | 8.10(5) | -0.07(3) | -0.15(6) | 11.58(8) |
| 9.8 | 0.37(5) | -0.07(3) | -0.11(5) | 12.34(7) |
| 13.6 | -11.39(4) | -0.08(2) | -0.06(5) | 13.62(7) |
| 20 | -26.83(4) | -0.09(2) | -0.08(4) | 15.53(6) |
| | GGA+U, $U = 7$ eV, $J_H = 1$ eV | | | |
| 0.6 | 48.46(10) | -0.09(5) | -0.41(11) | 10.69(16) |
| 3.2 | 31.40(9) | -0.074(5) | -0.23(10) | 12.13(14) |
| 4.8 | 22.67(7) | -0.07(4) | -0.13(8) | 12.93(11) |
| 9.8 | 0.42(7) | -0.06(4) | 0.00(8) | 15.20(11) |
| 13.6 | -13.14(7) | -0.07(4) | 0.05(8) | 16.78(11) |
| 20 | -31.05(6) | -0.08(3) | -0.01(6) | 19.14(10) |

that in these materials the angle is actually rather far from 90° is usually swept under the rug). This coupling is proportional to $t_{pd}^4 J_O / \Delta_{pd}^4$ (where $t_{pd}$ is the hopping and $\Delta_{pd}$ is the charge transfer energy). Note that the usual antiferromagnetic superexchange is proportional to $t_{pd}^4 / \Delta_{pd}^2 U$, which is larger by a factor $\Delta_{pd}^2 / U J_O$ for $\Delta_{pd}^2 / U J_O > 1$. Thus, near-cancellation of the assisted hopping via different O orbitals is a necessary requirement for ferromagnetic superexchange. There are two ways to look at this situation. One can assume a coordinate system where $\tilde{x}$ and $\tilde{y}$ are along the Cu-O and O-Cu bonds in the trimer, in which case the "left" Cu electron can only hop to the O $p_{\tilde{x}}$ orbital, and the "right" one only to the O $p_{\tilde{y}}$ one. Since these are mutually orthogonal, there is no net assisted Cu-Cu hopping. Another way is to assume the system where $x$ is along the Cu-Cu bond and $y$ is perpendicular to it. In this case, there is assisted hopping *via* each of the orbitals, but the two contributions are equal in magnitude and opposite in sign, so they cancel out. As the angle becomes larger than 90°, the Cu-$p_x$-Cu hopping becomes larger than Cu-$p_y$-Cu one and cancellation becomes incomplete. In typical transition metal oxides if the angle is larger than $\approx 95°$ this effect is significant.

However, our calculations as well as the experimental results, indicate that the crossover from the antiferromagnetic to ferromagnetic $J_1$ in these materials occurs at the bond angle of 108°−109°. Moreover, as discussed in Ref. [8] and confirmed in our calculations, shifting H away from O makes the interaction at a given angle less ferromagnetic, suggesting an important role of H orbitals.

In the literature one can find two seemingly contradicting predictions regarding adding to the standard 4-orbitals (two Cu $d$ and two O $d$) another bridging orbital. On one hand, Geertsma and Khomskii considered a $\phi = 90°$ case [10] when O $p$ orbitals are bound to another orbital, such as in our case hydrogen, in their case germanium, or, in the simplest case with the same symmetry, to O $s$. In that case the $p_y$ orbital couples with H, but $p_x$ does not. As a result, the cancellation described above is incomplete and there is some residual antiferromagnetic superexchange, which can either overcome the weak ferromagnetic coupling, as in Ref. [10], or considerably reduce it. Furthermore, this theory predicts that at $\phi = 90°$ moving H toward O must enhance the tendency to antiferromagnetism.

On the other hand, Hay *et al.* [11] considered a similar problem (using O $s$ and the additional bridging orbital, which is not important) as a function of $\phi$ and concluded that the full cancellation between all Cu-O-Cu hopping processes occurs not at $\phi = 90°$, but at a larger $\phi_c$ [for the Cu$_2$(OH)$_2$Cl$_4^{2-}$ radical they found numerically $\phi_c \approx 108°$, but the actual value must be system-dependent, and, in particular, may be smaller for our materials]. This is not in contradiction with Ref. [10]; indeed Hay *et al.* explicitly emphasize that it is only at $\phi_c$ that the antiferromagnetic exchange completely cancels out; for both larger *and* smaller angles there is residual antiferromagnetic coupling, in

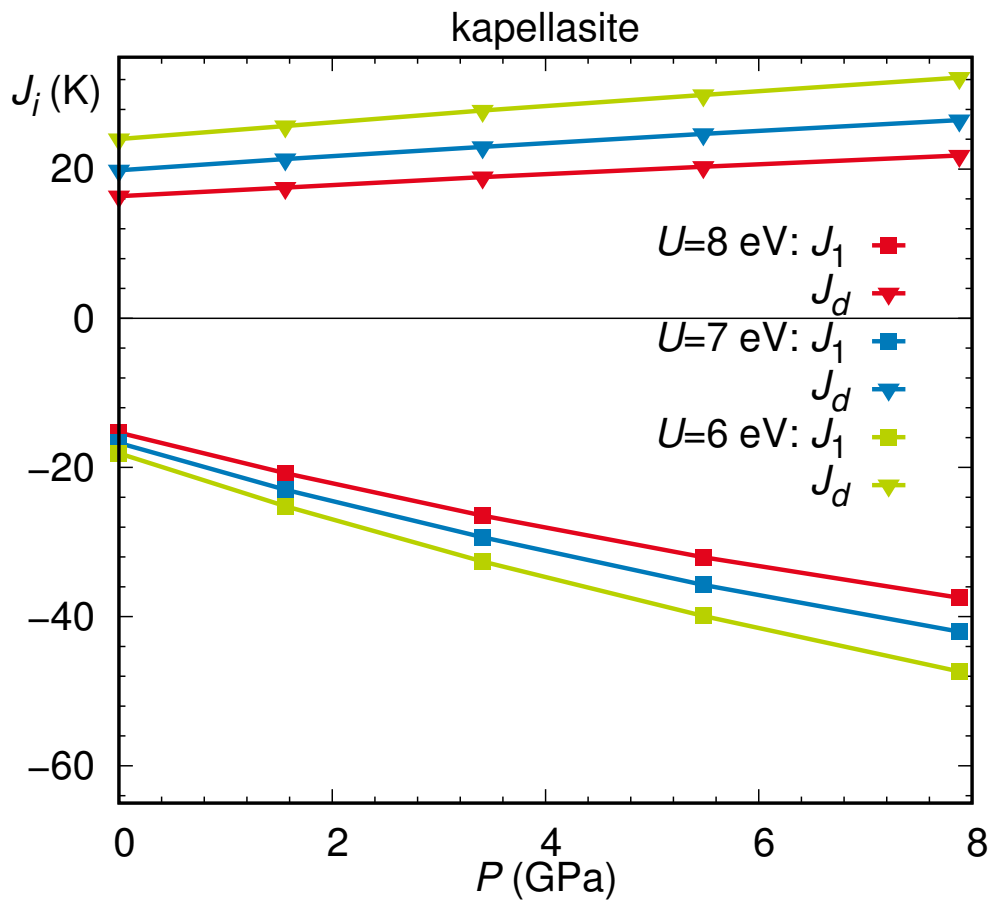

FIG. 1: **Exchange couplings of kapellasite.** The two significant Heisenberg Hamiltonian parameters for kapellasite are given as function of applied pressure and for three different values of the onsite interaction strength $U$.

accord with the Geertsma and Khomskii theory. However, at $\phi = 90°$ the Cu-$p_y$-Cu hopping is the stronger the shorter the O-H bond is, and to restore the balance one needs to enhance Cu-$p_x$-Cu hopping by flattening $\phi$. So, at least qualitatively (quantitative analysis is underway and will be published elsewhere), Hay *et al.* theory seems to explain why the large bond angles in HD and KL nontheless generate net ferromagnetic coupling.

Yet another paradox manifests itself in comparing the $J_d$ values for HD and KL. It is generally believed that this exchange is provided by hopping from Cu (*via* O) to Zn or Mg. However, as discussed below, this may be not the case. Indeed, Cu-(Zn,Mg) hopping can proceed either to the $s$ orbitals of Zn or Mg, or to $d$ orbitals of Zn. The $5s$ orbitals of Zn are considerably more diffuse than the $3s$ orbitals of Mg, and much closer to the Fermi level. Since $J_d$ in the simple Hubbard theory should be proportional to the square of the overlap integrals and inversely proportional to the energy separation, we estimate that $J_d$ in HD should be nearly an order of magnitude smaller than in KL, even without taking into account an additional hopping channel *via* $3d$ orbitals of Zn. In the experiment, as well as in the calculations, the difference is less than a factor of two. A closer look, however, shows that this naive reasoning doesn't apply here. Indeed, an inspection of the crystal structure shows that the Cu-O$_2$-(Zn,Mg)-O$_2$-Cu multimer is planar, as shown in Fig. 4. The bond angle is about 100° considerably closer to 90° that the Cu-O-Cu angle. Had it been exactly 90°, the hopping from the $x^2 - y^2$ Cu orbital, where the actual hole resides, to the (Zn,Mg) $s$ orbital would be zero by parity (see Fig. 4(b)). For 100°, it should be strongly suppressed. For Zn, there is an additional option of hopping via $3d$ orbitals. However, as Fig. 4(c) shows, $dp\sigma - pd\sigma$ hopping is again forbidden by symmetry. The $dp\sigma - pd\pi$ hopping is allowed, but the two oxygens are antiphase and cancel each other (see Fig. 4(d)).

Keeping in mind the H effect, one may conjecture that 100° may be rather close to full cancellation. Our numerical results provide for the effective (after integrating out O) Cu($d$)−Zn($d$) hopping a value of about $\tau \approx 40$ meV, which is nearly an order of magnitude smaller that for the Cu-Cu one. Thus, we conclude that the mechanism providing for sizeable $J_d$ in HD and KL, as compared to herbertsmithite, is more complex than it has been anticipated so




\begin{thebibliography}{58}%
\makeatletter
\providecommand \@ifxundefined [1]{%
 \@ifx{#1\undefined}
}%
\providecommand \@ifnum [1]{%
 \ifnum #1\expandafter \@firstoftwo
 \else \expandafter \@secondoftwo
 \fi
}%
\providecommand \@ifx [1]{%
 \ifx #1\expandafter \@firstoftwo
 \else \expandafter \@secondoftwo
 \fi
}%
\providecommand \natexlab [1]{#1}%
\providecommand \enquote  [1]{``#1''}%
\providecommand \bibnamefont  [1]{#1}%
\providecommand \bibfnamefont [1]{#1}%
\providecommand \citenamefont [1]{#1}%
\providecommand \href@noop [0]{\@secondoftwo}%
\providecommand \href [0]{\begingroup \@sanitize@url \@href}%
\providecommand \@href[1]{\@@startlink{#1}\@@href}%
\providecommand \@@href[1]{\endgroup#1\@@endlink}%
\providecommand \@sanitize@url [0]{\catcode `\\12\catcode `\$12\catcode
  `\&12\catcode `\#12\catcode `\^12\catcode `\_12\catcode `\%12\relax}%
\providecommand \@@startlink[1]{}%
\providecommand \@@endlink[0]{}%
\providecommand \url  [0]{\begingroup\@sanitize@url \@url }%
\providecommand \@url [1]{\endgroup\@href {#1}{\urlprefix }}%
\providecommand \urlprefix  [0]{URL }%
\providecommand \Eprint [0]{\href }%
\providecommand \doibase [0]{http://dx.doi.org/}%
\providecommand \selectlanguage [0]{\@gobble}%
\providecommand \bibinfo  [0]{\@secondoftwo}%
\providecommand \bibfield  [0]{\@secondoftwo}%
\providecommand \translation [1]{[#1]}%
\providecommand \BibitemOpen [0]{}%
\providecommand \bibitemStop [0]{}%
\providecommand \bibitemNoStop [0]{.\EOS\space}%
\providecommand \EOS [0]{\spacefactor3000\relax}%
\providecommand \BibitemShut  [1]{\csname bibitem#1\endcsname}%
\let\auto@bib@innerbib\@empty
%</preamble>
\bibitem [{\citenamefont {Pomeranchuk}(1941)}]{Pomeranchuk-1941}%
  \BibitemOpen
  \bibfield  {author} {\bibinfo {author} {\bibfnamefont {I.}~\bibnamefont
  {Pomeranchuk}},\ }\href@noop {} {\bibfield  {journal} {\bibinfo  {journal}
  {Zh. Eksp. Teor. Fiz.}\ }\textbf {\bibinfo {volume} {11}},\ \bibinfo {pages}
  {226} (\bibinfo {year} {1941})}\BibitemShut {NoStop}%
\bibitem [{\citenamefont {Balents}(2010)}]{Balents-2010}%
  \BibitemOpen
  \bibfield  {author} {\bibinfo {author} {\bibfnamefont {L.}~\bibnamefont
  {Balents}},\ }\href {http://dx.doi.org/10.1038/nature08917} {\bibfield
  {journal} {\bibinfo  {journal} {Nature (London)}\ }\textbf {\bibinfo {volume}
  {464}},\ \bibinfo {pages} {199} (\bibinfo {year} {2010})}\BibitemShut
  {NoStop}%
\bibitem [{\citenamefont {Singh}\ and\ \citenamefont
  {Huse}(2007)}]{Singh-2007}%
  \BibitemOpen
  \bibfield  {author} {\bibinfo {author} {\bibfnamefont {R.~R.~P.}\
  \bibnamefont {Singh}}\ and\ \bibinfo {author} {\bibfnamefont {D.~A.}\
  \bibnamefont {Huse}},\ }\href {\doibase 10.1103/PhysRevB.76.180407}
  {\bibfield  {journal} {\bibinfo  {journal} {Phys. Rev. B}\ }\textbf {\bibinfo
  {volume} {76}},\ \bibinfo {pages} {180407} (\bibinfo {year}
  {2007})}\BibitemShut {NoStop}%
\bibitem [{\citenamefont {Ran}\ \emph {et~al.}(2007)\citenamefont {Ran},
  \citenamefont {Hermele}, \citenamefont {Lee},\ and\ \citenamefont
  {Wen}}]{Ran-2007}%
  \BibitemOpen
  \bibfield  {author} {\bibinfo {author} {\bibfnamefont {Y.}~\bibnamefont
  {Ran}}, \bibinfo {author} {\bibfnamefont {M.}~\bibnamefont {Hermele}},
  \bibinfo {author} {\bibfnamefont {P.~A.}\ \bibnamefont {Lee}}, \ and\
  \bibinfo {author} {\bibfnamefont {X.-G.}\ \bibnamefont {Wen}},\ }\href
  {\doibase 10.1103/PhysRevLett.98.117205} {\bibfield  {journal} {\bibinfo
  {journal} {Phys. Rev. Lett.}\ }\textbf {\bibinfo {volume} {98}},\ \bibinfo
  {pages} {117205} (\bibinfo {year} {2007})}\BibitemShut {NoStop}%
\bibitem [{\citenamefont {Nakano}\ and\ \citenamefont
  {Sakai}(2011)}]{Nakano-2011}%
  \BibitemOpen
  \bibfield  {author} {\bibinfo {author} {\bibfnamefont {H.}~\bibnamefont
  {Nakano}}\ and\ \bibinfo {author} {\bibfnamefont {T.}~\bibnamefont {Sakai}},\
  }\href {\doibase 10.1143/JPSJ.80.053704} {\bibfield  {journal} {\bibinfo
  {journal} {J. Phys. Soc. Jpn.}\ }\textbf {\bibinfo {volume} {80}},\ \bibinfo
  {pages} {053704} (\bibinfo {year} {2011})}\BibitemShut {NoStop}%
\bibitem [{\citenamefont {L\"auchli}\ \emph {et~al.}(2011)\citenamefont
  {L\"auchli}, \citenamefont {Sudan},\ and\ \citenamefont
  {S\o{}rensen}}]{Lauchli-2011}%
  \BibitemOpen
  \bibfield  {author} {\bibinfo {author} {\bibfnamefont {A.~M.}\ \bibnamefont
  {L\"auchli}}, \bibinfo {author} {\bibfnamefont {J.}~\bibnamefont {Sudan}}, \
  and\ \bibinfo {author} {\bibfnamefont {E.~S.}\ \bibnamefont {S\o{}rensen}},\
  }\href {\doibase 10.1103/PhysRevB.83.212401} {\bibfield  {journal} {\bibinfo
  {journal} {Phys. Rev. B}\ }\textbf {\bibinfo {volume} {83}},\ \bibinfo
  {pages} {212401} (\bibinfo {year} {2011})}\BibitemShut {NoStop}%
\bibitem [{\citenamefont {Iqbal}\ \emph
  {et~al.}(2011{\natexlab{a}})\citenamefont {Iqbal}, \citenamefont {Becca},\
  and\ \citenamefont {Poilblanc}}]{Iqbal-2011a}%
  \BibitemOpen
  \bibfield  {author} {\bibinfo {author} {\bibfnamefont {Y.}~\bibnamefont
  {Iqbal}}, \bibinfo {author} {\bibfnamefont {F.}~\bibnamefont {Becca}}, \ and\
  \bibinfo {author} {\bibfnamefont {D.}~\bibnamefont {Poilblanc}},\ }\href
  {\doibase 10.1103/PhysRevB.83.100404} {\bibfield  {journal} {\bibinfo
  {journal} {Phys. Rev. B}\ }\textbf {\bibinfo {volume} {83}},\ \bibinfo
  {pages} {100404} (\bibinfo {year} {2011}{\natexlab{a}})}\BibitemShut
  {NoStop}%
\bibitem [{\citenamefont {Iqbal}\ \emph
  {et~al.}(2011{\natexlab{b}})\citenamefont {Iqbal}, \citenamefont {Becca},\
  and\ \citenamefont {Poilblanc}}]{Iqbal-2011b}%
  \BibitemOpen
  \bibfield  {author} {\bibinfo {author} {\bibfnamefont {Y.}~\bibnamefont
  {Iqbal}}, \bibinfo {author} {\bibfnamefont {F.}~\bibnamefont {Becca}}, \ and\
  \bibinfo {author} {\bibfnamefont {D.}~\bibnamefont {Poilblanc}},\ }\href
  {\doibase 10.1103/PhysRevB.84.020407} {\bibfield  {journal} {\bibinfo
  {journal} {Phys. Rev. B}\ }\textbf {\bibinfo {volume} {84}},\ \bibinfo
  {pages} {020407} (\bibinfo {year} {2011}{\natexlab{b}})}\BibitemShut
  {NoStop}%
\bibitem [{\citenamefont {Yan}\ \emph {et~al.}(2011)\citenamefont {Yan},
  \citenamefont {Huse},\ and\ \citenamefont {White}}]{Yan-2011}%
  \BibitemOpen
  \bibfield  {author} {\bibinfo {author} {\bibfnamefont {S.}~\bibnamefont
  {Yan}}, \bibinfo {author} {\bibfnamefont {D.~A.}\ \bibnamefont {Huse}}, \
  and\ \bibinfo {author} {\bibfnamefont {S.~R.}\ \bibnamefont {White}},\ }\href
  {\doibase 10.1126/science.1201080} {\bibfield  {journal} {\bibinfo  {journal}
  {Science}\ }\textbf {\bibinfo {volume} {332}},\ \bibinfo {pages} {1173}
  (\bibinfo {year} {2011})}\BibitemShut {NoStop}%
\bibitem [{\citenamefont {Depenbrock}\ \emph {et~al.}(2012)\citenamefont
  {Depenbrock}, \citenamefont {McCulloch},\ and\ \citenamefont
  {Schollw\"ock}}]{Depenbrock-2012}%
  \BibitemOpen
  \bibfield  {author} {\bibinfo {author} {\bibfnamefont {S.}~\bibnamefont
  {Depenbrock}}, \bibinfo {author} {\bibfnamefont {I.~P.}\ \bibnamefont
  {McCulloch}}, \ and\ \bibinfo {author} {\bibfnamefont {U.}~\bibnamefont
  {Schollw\"ock}},\ }\href {\doibase 10.1103/PhysRevLett.109.067201} {\bibfield
   {journal} {\bibinfo  {journal} {Phys. Rev. Lett.}\ }\textbf {\bibinfo
  {volume} {109}},\ \bibinfo {pages} {067201} (\bibinfo {year}
  {2012})}\BibitemShut {NoStop}%
\bibitem [{\citenamefont {Jiang}\ \emph {et~al.}(2012)\citenamefont {Jiang},
  \citenamefont {Wang},\ and\ \citenamefont {Balents}}]{Jiang-2012}%
  \BibitemOpen
  \bibfield  {author} {\bibinfo {author} {\bibfnamefont {H.-C.}\ \bibnamefont
  {Jiang}}, \bibinfo {author} {\bibfnamefont {Z.}~\bibnamefont {Wang}}, \ and\
  \bibinfo {author} {\bibfnamefont {L.}~\bibnamefont {Balents}},\ }\href
  {http://dx.doi.org/10.1038/nphys2465} {\bibfield  {journal} {\bibinfo
  {journal} {Nat. Phys.}\ }\textbf {\bibinfo {volume} {8}},\ \bibinfo {pages}
  {902} (\bibinfo {year} {2012})}\BibitemShut {NoStop}%
\bibitem [{\citenamefont {Capponi}\ \emph {et~al.}(2013)\citenamefont
  {Capponi}, \citenamefont {Chandra}, \citenamefont {Auerbach},\ and\
  \citenamefont {Weinstein}}]{Capponi-2013}%
  \BibitemOpen
  \bibfield  {author} {\bibinfo {author} {\bibfnamefont {S.}~\bibnamefont
  {Capponi}}, \bibinfo {author} {\bibfnamefont {V.~R.}\ \bibnamefont
  {Chandra}}, \bibinfo {author} {\bibfnamefont {A.}~\bibnamefont {Auerbach}}, \
  and\ \bibinfo {author} {\bibfnamefont {M.}~\bibnamefont {Weinstein}},\ }\href
  {\doibase 10.1103/PhysRevB.87.161118} {\bibfield  {journal} {\bibinfo
  {journal} {Phys. Rev. B}\ }\textbf {\bibinfo {volume} {87}},\ \bibinfo
  {pages} {161118} (\bibinfo {year} {2013})}\BibitemShut {NoStop}%
\bibitem [{\citenamefont {Iqbal}\ \emph {et~al.}(2013)\citenamefont {Iqbal},
  \citenamefont {Becca}, \citenamefont {Sorella},\ and\ \citenamefont
  {Poilblanc}}]{Iqbal-2013}%
  \BibitemOpen
  \bibfield  {author} {\bibinfo {author} {\bibfnamefont {Y.}~\bibnamefont
  {Iqbal}}, \bibinfo {author} {\bibfnamefont {F.}~\bibnamefont {Becca}},
  \bibinfo {author} {\bibfnamefont {S.}~\bibnamefont {Sorella}}, \ and\
  \bibinfo {author} {\bibfnamefont {D.}~\bibnamefont {Poilblanc}},\ }\href
  {\doibase 10.1103/PhysRevB.87.060405} {\bibfield  {journal} {\bibinfo
  {journal} {Phys. Rev. B}\ }\textbf {\bibinfo {volume} {87}},\ \bibinfo
  {pages} {060405} (\bibinfo {year} {2013})}\BibitemShut {NoStop}%
\bibitem [{\citenamefont {Xie}\ \emph {et~al.}(2014)\citenamefont {Xie},
  \citenamefont {Chen}, \citenamefont {Yu}, \citenamefont {Kong}, \citenamefont
  {Normand},\ and\ \citenamefont {Xiang}}]{Xie-2014}%
  \BibitemOpen
  \bibfield  {author} {\bibinfo {author} {\bibfnamefont {Y.}~\bibnamefont
  {Xie}, \bibfnamefont {Z.}}, \bibinfo {author} {\bibfnamefont
  {J.}~\bibnamefont {Chen}}, \bibinfo {author} {\bibfnamefont {F.}~\bibnamefont
  {Yu}, \bibfnamefont {J.}}, \bibinfo {author} {\bibfnamefont {X.}~\bibnamefont
  {Kong}}, \bibinfo {author} {\bibfnamefont {B.}~\bibnamefont {Normand}}, \
  and\ \bibinfo {author} {\bibfnamefont {T.}~\bibnamefont {Xiang}},\ }\href
  {\doibase 10.1103/PhysRevX.4.011025} {\bibfield  {journal} {\bibinfo
  {journal} {Phys. Rev. X}\ }\textbf {\bibinfo {volume} {4}},\ \bibinfo {pages}
  {011025} (\bibinfo {year} {2014})}\BibitemShut {NoStop}%
\bibitem [{\citenamefont {Punk}\ \emph {et~al.}(2014)\citenamefont {Punk},
  \citenamefont {Chowdhury},\ and\ \citenamefont {Sachdev}}]{Punk-2014}%
  \BibitemOpen
  \bibfield  {author} {\bibinfo {author} {\bibfnamefont {M.}~\bibnamefont
  {Punk}}, \bibinfo {author} {\bibfnamefont {D.}~\bibnamefont {Chowdhury}}, \
  and\ \bibinfo {author} {\bibfnamefont {S.}~\bibnamefont {Sachdev}},\ }\href
  {http://dx.doi.org/10.1038/nphys2887} {\bibfield  {journal} {\bibinfo
  {journal} {Nat. Phys.}\ }\textbf {\bibinfo {volume} {10}},\ \bibinfo {pages}
  {289} (\bibinfo {year} {2014})}\BibitemShut {NoStop}%
\bibitem [{\citenamefont {Iqbal}\ \emph {et~al.}(2014)\citenamefont {Iqbal},
  \citenamefont {Poilblanc},\ and\ \citenamefont {Becca}}]{Iqbal-2014}%
  \BibitemOpen
  \bibfield  {author} {\bibinfo {author} {\bibfnamefont {Y.}~\bibnamefont
  {Iqbal}}, \bibinfo {author} {\bibfnamefont {D.}~\bibnamefont {Poilblanc}}, \
  and\ \bibinfo {author} {\bibfnamefont {F.}~\bibnamefont {Becca}},\ }\href
  {\doibase 10.1103/PhysRevB.89.020407} {\bibfield  {journal} {\bibinfo
  {journal} {Phys. Rev. B}\ }\textbf {\bibinfo {volume} {89}},\ \bibinfo
  {pages} {020407} (\bibinfo {year} {2014})}\BibitemShut {NoStop}%
\bibitem [{\citenamefont {Nandkishore}\ \emph {et~al.}(2013)\citenamefont
  {Nandkishore}, \citenamefont {Maciejko}, \citenamefont {Huse},\ and\
  \citenamefont {Sondhi}}]{Nandkishore-2013}%
  \BibitemOpen
  \bibfield  {author} {\bibinfo {author} {\bibfnamefont {R.}~\bibnamefont
  {Nandkishore}}, \bibinfo {author} {\bibfnamefont {J.}~\bibnamefont
  {Maciejko}}, \bibinfo {author} {\bibfnamefont {D.~A.}\ \bibnamefont {Huse}},
  \ and\ \bibinfo {author} {\bibfnamefont {S.~L.}\ \bibnamefont {Sondhi}},\
  }\href {\doibase 10.1103/PhysRevB.87.174511} {\bibfield  {journal} {\bibinfo
  {journal} {Phys. Rev. B}\ }\textbf {\bibinfo {volume} {87}},\ \bibinfo
  {pages} {174511} (\bibinfo {year} {2013})}\BibitemShut {NoStop}%
\bibitem [{\citenamefont {Mazin}\ \emph {et~al.}(2014)\citenamefont {Mazin},
  \citenamefont {Jeschke}, \citenamefont {Lechermann}, \citenamefont {Lee},
  \citenamefont {Fink}, \citenamefont {Thomale},\ and\ \citenamefont
  {Valent{\'\i}}}]{Mazin-2014}%
  \BibitemOpen
  \bibfield  {author} {\bibinfo {author} {\bibfnamefont {I.~I.}\ \bibnamefont
  {Mazin}}, \bibinfo {author} {\bibfnamefont {H.~O.}\ \bibnamefont {Jeschke}},
  \bibinfo {author} {\bibfnamefont {F.}~\bibnamefont {Lechermann}}, \bibinfo
  {author} {\bibfnamefont {H.}~\bibnamefont {Lee}}, \bibinfo {author}
  {\bibfnamefont {M.}~\bibnamefont {Fink}}, \bibinfo {author} {\bibfnamefont
  {R.}~\bibnamefont {Thomale}}, \ and\ \bibinfo {author} {\bibfnamefont
  {R.}~\bibnamefont {Valent{\'\i}}},\ }\href
  {http://dx.doi.org/10.1038/ncomms5261} {\bibfield  {journal} {\bibinfo
  {journal} {Nat. Commun.}\ }\textbf {\bibinfo {volume} {5}} (\bibinfo {year}
  {2014})}\BibitemShut {NoStop}%
\bibitem [{\citenamefont {Shores}\ \emph {et~al.}(2005)\citenamefont {Shores},
  \citenamefont {Nytko}, \citenamefont {Bartlett},\ and\ \citenamefont
  {Nocera}}]{Shores-2005}%
  \BibitemOpen
  \bibfield  {author} {\bibinfo {author} {\bibfnamefont {M.~P.}\ \bibnamefont
  {Shores}}, \bibinfo {author} {\bibfnamefont {E.~A.}\ \bibnamefont {Nytko}},
  \bibinfo {author} {\bibfnamefont {B.~M.}\ \bibnamefont {Bartlett}}, \ and\
  \bibinfo {author} {\bibfnamefont {D.~G.}\ \bibnamefont {Nocera}},\ }\href
  {\doibase 10.1021/ja053891p} {\bibfield  {journal} {\bibinfo  {journal} {J.
  Am. Chem. Soc.}\ }\textbf {\bibinfo {volume} {127}},\ \bibinfo {pages}
  {13462} (\bibinfo {year} {2005})}\BibitemShut {NoStop}%
\bibitem [{\citenamefont {Han}\ \emph {et~al.}(2012{\natexlab{a}})\citenamefont
  {Han}, \citenamefont {Helton}, \citenamefont {Chu}, \citenamefont {Nocera},
  \citenamefont {Rodriguez-Rivera}, \citenamefont {Broholm},\ and\
  \citenamefont {Lee}}]{Han-2012a}%
  \BibitemOpen
  \bibfield  {author} {\bibinfo {author} {\bibfnamefont {T.-H.}\ \bibnamefont
  {Han}}, \bibinfo {author} {\bibfnamefont {J.~S.}\ \bibnamefont {Helton}},
  \bibinfo {author} {\bibfnamefont {S.}~\bibnamefont {Chu}}, \bibinfo {author}
  {\bibfnamefont {D.~G.}\ \bibnamefont {Nocera}}, \bibinfo {author}
  {\bibfnamefont {J.~A.}\ \bibnamefont {Rodriguez-Rivera}}, \bibinfo {author}
  {\bibfnamefont {C.}~\bibnamefont {Broholm}}, \ and\ \bibinfo {author}
  {\bibfnamefont {Y.~S.}\ \bibnamefont {Lee}},\ }\href
  {http://dx.doi.org/10.1038/nature11659} {\bibfield  {journal} {\bibinfo
  {journal} {Nature (London)}\ }\textbf {\bibinfo {volume} {492}},\ \bibinfo
  {pages} {406} (\bibinfo {year} {2012}{\natexlab{a}})}\BibitemShut {NoStop}%
\bibitem [{\citenamefont {Krause}\ \emph {et~al.}(2006)\citenamefont {Krause},
  \citenamefont {Bernhardt}, \citenamefont {Braithwaite}, \citenamefont
  {Kolitsch},\ and\ \citenamefont {Pritchard}}]{Krause-2006}%
  \BibitemOpen
  \bibfield  {author} {\bibinfo {author} {\bibfnamefont {W.}~\bibnamefont
  {Krause}}, \bibinfo {author} {\bibfnamefont {H.-J.}\ \bibnamefont
  {Bernhardt}}, \bibinfo {author} {\bibfnamefont {R.~S.~W.}\ \bibnamefont
  {Braithwaite}}, \bibinfo {author} {\bibfnamefont {U.}~\bibnamefont
  {Kolitsch}}, \ and\ \bibinfo {author} {\bibfnamefont {R.}~\bibnamefont
  {Pritchard}},\ }\href {\doibase 10.1180/0026461067030336} {\bibfield
  {journal} {\bibinfo  {journal} {Mineral. Mag.}\ }\textbf {\bibinfo {volume}
  {70}},\ \bibinfo {pages} {329} (\bibinfo {year} {2006})}\BibitemShut
  {NoStop}%
\bibitem [{\citenamefont {Colman}\ \emph {et~al.}(2008)\citenamefont {Colman},
  \citenamefont {Ritter},\ and\ \citenamefont {Wills}}]{Colman-2008}%
  \BibitemOpen
  \bibfield  {author} {\bibinfo {author} {\bibfnamefont {R.~H.}\ \bibnamefont
  {Colman}}, \bibinfo {author} {\bibfnamefont {C.}~\bibnamefont {Ritter}}, \
  and\ \bibinfo {author} {\bibfnamefont {A.~S.}\ \bibnamefont {Wills}},\ }\href
  {\doibase 10.1021/cm802060n} {\bibfield  {journal} {\bibinfo  {journal}
  {Chem. Mater.}\ }\textbf {\bibinfo {volume} {20}},\ \bibinfo {pages} {6897}
  (\bibinfo {year} {2008})}\BibitemShut {NoStop}%
\bibitem [{\citenamefont {Malcherek}\ and\ \citenamefont
  {Schl\"uter}(2007)}]{Malcherek-2007}%
  \BibitemOpen
  \bibfield  {author} {\bibinfo {author} {\bibfnamefont {T.}~\bibnamefont
  {Malcherek}}\ and\ \bibinfo {author} {\bibfnamefont {J.}~\bibnamefont
  {Schl\"uter}},\ }\href {\doibase 10.1107/S0108768106053870} {\bibfield
  {journal} {\bibinfo  {journal} {Acta Crystallogr., Sect. B: Struct. Sci.}\
  }\textbf {\bibinfo {volume} {63}},\ \bibinfo {pages} {157} (\bibinfo {year}
  {2007})}\BibitemShut {NoStop}%
\bibitem [{\citenamefont {Schl{\"u}ter}\ and\ \citenamefont
  {Malcherek}(2007)}]{Schlueter-2007}%
  \BibitemOpen
  \bibfield  {author} {\bibinfo {author} {\bibfnamefont {J.}~\bibnamefont
  {Schl{\"u}ter}}\ and\ \bibinfo {author} {\bibfnamefont {T.}~\bibnamefont
  {Malcherek}},\ }\href {http://dx.doi.org/10.1127/0077-7757/2007/0086}
  {\bibfield  {journal} {\bibinfo  {journal} {Neues Jahrbuch f{\"u}r
  Mineralogie - Abhandlungen}\ }\textbf {\bibinfo {volume} {184}},\ \bibinfo
  {pages} {39} (\bibinfo {year} {2007})}\BibitemShut {NoStop}%
\bibitem [{\citenamefont {Colman}\ \emph {et~al.}(2010)\citenamefont {Colman},
  \citenamefont {Sinclair},\ and\ \citenamefont {Wills}}]{Colman-2010}%
  \BibitemOpen
  \bibfield  {author} {\bibinfo {author} {\bibfnamefont {R.}~\bibnamefont
  {Colman}}, \bibinfo {author} {\bibfnamefont {A.}~\bibnamefont {Sinclair}}, \
  and\ \bibinfo {author} {\bibfnamefont {A.}~\bibnamefont {Wills}},\ }\href
  {\doibase 10.1021/cm101594c} {\bibfield  {journal} {\bibinfo  {journal}
  {Chem. Mater.}\ }\textbf {\bibinfo {volume} {22}},\ \bibinfo {pages} {5774}
  (\bibinfo {year} {2010})}\BibitemShut {NoStop}%
\bibitem [{\citenamefont {Chu}(2011)}]{Chu-2011}%
  \BibitemOpen
  \bibfield  {author} {\bibinfo {author} {\bibfnamefont {S.}~\bibnamefont
  {Chu}},\ }\href {http://stacks.iop.org/1742-6596/273/i=1/a=012123} {\bibfield
   {journal} {\bibinfo  {journal} {J. Phys.: Conf. Ser.}\ }\textbf {\bibinfo
  {volume} {273}},\ \bibinfo {pages} {012123} (\bibinfo {year}
  {2011})}\BibitemShut {NoStop}%
\bibitem [{\citenamefont {Boldrin}\ \emph {et~al.}(2015)\citenamefont
  {Boldrin}, \citenamefont {F\aa{}k}, \citenamefont {Enderle}, \citenamefont
  {Bieri}, \citenamefont {Ollivier}, \citenamefont {Rols}, \citenamefont
  {Manuel},\ and\ \citenamefont {Wills}}]{Boldrin-2015}%
  \BibitemOpen
  \bibfield  {author} {\bibinfo {author} {\bibfnamefont {D.}~\bibnamefont
  {Boldrin}}, \bibinfo {author} {\bibfnamefont {B.}~\bibnamefont {F\aa{}k}},
  \bibinfo {author} {\bibfnamefont {M.}~\bibnamefont {Enderle}}, \bibinfo
  {author} {\bibfnamefont {S.}~\bibnamefont {Bieri}}, \bibinfo {author}
  {\bibfnamefont {J.}~\bibnamefont {Ollivier}}, \bibinfo {author}
  {\bibfnamefont {S.}~\bibnamefont {Rols}}, \bibinfo {author} {\bibfnamefont
  {P.}~\bibnamefont {Manuel}}, \ and\ \bibinfo {author} {\bibfnamefont {A.~S.}\
  \bibnamefont {Wills}},\ }\href {\doibase 10.1103/PhysRevB.91.220408}
  {\bibfield  {journal} {\bibinfo  {journal} {Phys. Rev. B}\ }\textbf {\bibinfo
  {volume} {91}},\ \bibinfo {pages} {220408} (\bibinfo {year}
  {2015})}\BibitemShut {NoStop}%
\bibitem [{\citenamefont {Reuther}\ and\ \citenamefont
  {W\"olfle}(2010)}]{Reuther-2010}%
  \BibitemOpen
  \bibfield  {author} {\bibinfo {author} {\bibfnamefont {J.}~\bibnamefont
  {Reuther}}\ and\ \bibinfo {author} {\bibfnamefont {P.}~\bibnamefont
  {W\"olfle}},\ }\href {\doibase 10.1103/PhysRevB.81.144410} {\bibfield
  {journal} {\bibinfo  {journal} {Phys. Rev. B}\ }\textbf {\bibinfo {volume}
  {81}},\ \bibinfo {pages} {144410} (\bibinfo {year} {2010})}\BibitemShut
  {NoStop}%
\bibitem [{\citenamefont {Reuther}\ and\ \citenamefont
  {Thomale}(2011)}]{Reuther-2011a}%
  \BibitemOpen
  \bibfield  {author} {\bibinfo {author} {\bibfnamefont {J.}~\bibnamefont
  {Reuther}}\ and\ \bibinfo {author} {\bibfnamefont {R.}~\bibnamefont
  {Thomale}},\ }\href {\doibase 10.1103/PhysRevB.83.024402} {\bibfield
  {journal} {\bibinfo  {journal} {Phys. Rev. B}\ }\textbf {\bibinfo {volume}
  {83}},\ \bibinfo {pages} {024402} (\bibinfo {year} {2011})}\BibitemShut
  {NoStop}%
\bibitem [{\citenamefont {Suttner}\ \emph {et~al.}(2014)\citenamefont
  {Suttner}, \citenamefont {Platt}, \citenamefont {Reuther},\ and\
  \citenamefont {Thomale}}]{Suttner-2014}%
  \BibitemOpen
  \bibfield  {author} {\bibinfo {author} {\bibfnamefont {R.}~\bibnamefont
  {Suttner}}, \bibinfo {author} {\bibfnamefont {C.}~\bibnamefont {Platt}},
  \bibinfo {author} {\bibfnamefont {J.}~\bibnamefont {Reuther}}, \ and\
  \bibinfo {author} {\bibfnamefont {R.}~\bibnamefont {Thomale}},\ }\href
  {\doibase 10.1103/PhysRevB.89.020408} {\bibfield  {journal} {\bibinfo
  {journal} {Phys. Rev. B}\ }\textbf {\bibinfo {volume} {89}},\ \bibinfo
  {pages} {020408} (\bibinfo {year} {2014})}\BibitemShut {NoStop}%
\bibitem [{\citenamefont {Han}\ \emph {et~al.}(2012{\natexlab{b}})\citenamefont
  {Han}, \citenamefont {Chu},\ and\ \citenamefont {Lee}}]{Han-2012b}%
  \BibitemOpen
  \bibfield  {author} {\bibinfo {author} {\bibfnamefont {T.}~\bibnamefont
  {Han}}, \bibinfo {author} {\bibfnamefont {S.}~\bibnamefont {Chu}}, \ and\
  \bibinfo {author} {\bibfnamefont {Y.~S.}\ \bibnamefont {Lee}},\ }\href
  {\doibase 10.1103/PhysRevLett.108.157202} {\bibfield  {journal} {\bibinfo
  {journal} {Phys. Rev. Lett.}\ }\textbf {\bibinfo {volume} {108}},\ \bibinfo
  {pages} {157202} (\bibinfo {year} {2012}{\natexlab{b}})}\BibitemShut
  {NoStop}%
\bibitem [{\citenamefont {Misguich}\ and\ \citenamefont
  {Sindzingre}(2007)}]{Misguich-2007}%
  \BibitemOpen
  \bibfield  {author} {\bibinfo {author} {\bibfnamefont {G.}~\bibnamefont
  {Misguich}}\ and\ \bibinfo {author} {\bibfnamefont {P.}~\bibnamefont
  {Sindzingre}},\ }\href {\doibase 10.1140/epjb/e2007-00301-6} {\bibfield
  {journal} {\bibinfo  {journal} {Eur. Phys. J. B}\ }\textbf {\bibinfo {volume}
  {59}},\ \bibinfo {pages} {305} (\bibinfo {year} {2007})}\BibitemShut
  {NoStop}%
\bibitem [{\citenamefont {Jeschke}\ \emph {et~al.}(2013)\citenamefont
  {Jeschke}, \citenamefont {Salvat-Pujol},\ and\ \citenamefont
  {Valent\'\i}}]{Valenti-2013}%
  \BibitemOpen
  \bibfield  {author} {\bibinfo {author} {\bibfnamefont {H.~O.}\ \bibnamefont
  {Jeschke}}, \bibinfo {author} {\bibfnamefont {F.}~\bibnamefont
  {Salvat-Pujol}}, \ and\ \bibinfo {author} {\bibfnamefont {R.}~\bibnamefont
  {Valent\'\i}},\ }\href {\doibase 10.1103/PhysRevB.88.075106} {\bibfield
  {journal} {\bibinfo  {journal} {Phys. Rev. B}\ }\textbf {\bibinfo {volume}
  {88}},\ \bibinfo {pages} {075106} (\bibinfo {year} {2013})}\BibitemShut
  {NoStop}%
\bibitem [{\citenamefont {Iqbal}\ \emph {et~al.}(2015)\citenamefont {Iqbal},
  \citenamefont {Poilblanc},\ and\ \citenamefont {Becca}}]{Iqbal-2015}%
  \BibitemOpen
  \bibfield  {author} {\bibinfo {author} {\bibfnamefont {Y.}~\bibnamefont
  {Iqbal}}, \bibinfo {author} {\bibfnamefont {D.}~\bibnamefont {Poilblanc}}, \
  and\ \bibinfo {author} {\bibfnamefont {F.}~\bibnamefont {Becca}},\ }\href
  {\doibase 10.1103/PhysRevB.91.020402} {\bibfield  {journal} {\bibinfo
  {journal} {Phys. Rev. B}\ }\textbf {\bibinfo {volume} {91}},\ \bibinfo
  {pages} {020402} (\bibinfo {year} {2015})}\BibitemShut {NoStop}%
\bibitem [{\citenamefont {Iqbal}\ \emph {et~al.}(2012)\citenamefont {Iqbal},
  \citenamefont {Becca},\ and\ \citenamefont {Poilblanc}}]{Iqbal-2012}%
  \BibitemOpen
  \bibfield  {author} {\bibinfo {author} {\bibfnamefont {Y.}~\bibnamefont
  {Iqbal}}, \bibinfo {author} {\bibfnamefont {F.}~\bibnamefont {Becca}}, \ and\
  \bibinfo {author} {\bibfnamefont {D.}~\bibnamefont {Poilblanc}},\ }\href
  {http://stacks.iop.org/1367-2630/14/i=11/a=115031} {\bibfield  {journal}
  {\bibinfo  {journal} {New J. Phys.}\ }\textbf {\bibinfo {volume} {14}},\
  \bibinfo {pages} {115031} (\bibinfo {year} {2012})}\BibitemShut {NoStop}%
\bibitem [{\citenamefont {Becca}\ \emph {et~al.}(2015)\citenamefont {Becca},
  \citenamefont {Hu}, \citenamefont {Iqbal}, \citenamefont {Parola},
  \citenamefont {Poilblanc},\ and\ \citenamefont {Sorella}}]{Becca-2015}%
  \BibitemOpen
  \bibfield  {author} {\bibinfo {author} {\bibfnamefont {F.}~\bibnamefont
  {Becca}}, \bibinfo {author} {\bibfnamefont {W.-J.}\ \bibnamefont {Hu}},
  \bibinfo {author} {\bibfnamefont {Y.}~\bibnamefont {Iqbal}}, \bibinfo
  {author} {\bibfnamefont {A.}~\bibnamefont {Parola}}, \bibinfo {author}
  {\bibfnamefont {D.}~\bibnamefont {Poilblanc}}, \ and\ \bibinfo {author}
  {\bibfnamefont {S.}~\bibnamefont {Sorella}},\ }\href
  {http://stacks.iop.org/1742-6596/640/i=1/a=012039} {\bibfield  {journal}
  {\bibinfo  {journal} {J. Phys.: Conf. Ser.}\ }\textbf {\bibinfo {volume}
  {640}},\ \bibinfo {pages} {012039} (\bibinfo {year} {2015})}\BibitemShut
  {NoStop}%
\bibitem [{\citenamefont {Mendels}\ \emph {et~al.}(2007)\citenamefont
  {Mendels}, \citenamefont {Bert}, \citenamefont {de~Vries}, \citenamefont
  {Olariu}, \citenamefont {Harrison}, \citenamefont {Duc}, \citenamefont
  {Trombe}, \citenamefont {Lord}, \citenamefont {Amato},\ and\ \citenamefont
  {Baines}}]{Mendels-2007}%
  \BibitemOpen
  \bibfield  {author} {\bibinfo {author} {\bibfnamefont {P.}~\bibnamefont
  {Mendels}}, \bibinfo {author} {\bibfnamefont {F.}~\bibnamefont {Bert}},
  \bibinfo {author} {\bibfnamefont {M.~A.}\ \bibnamefont {de~Vries}}, \bibinfo
  {author} {\bibfnamefont {A.}~\bibnamefont {Olariu}}, \bibinfo {author}
  {\bibfnamefont {A.}~\bibnamefont {Harrison}}, \bibinfo {author}
  {\bibfnamefont {F.}~\bibnamefont {Duc}}, \bibinfo {author} {\bibfnamefont
  {J.~C.}\ \bibnamefont {Trombe}}, \bibinfo {author} {\bibfnamefont {J.~S.}\
  \bibnamefont {Lord}}, \bibinfo {author} {\bibfnamefont {A.}~\bibnamefont
  {Amato}}, \ and\ \bibinfo {author} {\bibfnamefont {C.}~\bibnamefont
  {Baines}},\ }\href {\doibase 10.1103/PhysRevLett.98.077204} {\bibfield
  {journal} {\bibinfo  {journal} {Phys. Rev. Lett.}\ }\textbf {\bibinfo
  {volume} {98}},\ \bibinfo {pages} {077204} (\bibinfo {year}
  {2007})}\BibitemShut {NoStop}%
\bibitem [{\citenamefont {Fu}\ \emph {et~al.}(2015)\citenamefont {Fu},
  \citenamefont {Imai}, \citenamefont {Han},\ and\ \citenamefont
  {Lee}}]{Fu-2015}%
  \BibitemOpen
  \bibfield  {author} {\bibinfo {author} {\bibfnamefont {M.}~\bibnamefont
  {Fu}}, \bibinfo {author} {\bibfnamefont {T.}~\bibnamefont {Imai}}, \bibinfo
  {author} {\bibfnamefont {T.-H.}\ \bibnamefont {Han}}, \ and\ \bibinfo
  {author} {\bibfnamefont {Y.~S.}\ \bibnamefont {Lee}},\ }\href {\doibase
  10.1126/science.aab2120} {\bibfield  {journal} {\bibinfo  {journal}
  {Science}\ }\textbf {\bibinfo {volume} {350}},\ \bibinfo {pages} {655}
  (\bibinfo {year} {2015})}\BibitemShut {NoStop}%
\bibitem [{Sup()}]{Supp}%
  \BibitemOpen
  \href@noop {} {\bibinfo  {journal} {See Supplemental Material at the end of
  the main article for a discussion on the origin of exchange parameters and
  details of {\it ab initio} calculations}\ }\BibitemShut {NoStop}%
\bibitem [{\citenamefont {Bernu}\ \emph {et~al.}(2013)\citenamefont {Bernu},
  \citenamefont {Lhuillier}, \citenamefont {Kermarrec}, \citenamefont {Bert},
  \citenamefont {Mendels}, \citenamefont {Colman},\ and\ \citenamefont
  {Wills}}]{Bernu-2013}%
  \BibitemOpen
\bibfield  {journal} {  }\bibfield  {author} {\bibinfo {author} {\bibfnamefont
  {B.}~\bibnamefont {Bernu}}, \bibinfo {author} {\bibfnamefont
  {C.}~\bibnamefont {Lhuillier}}, \bibinfo {author} {\bibfnamefont
  {E.}~\bibnamefont {Kermarrec}}, \bibinfo {author} {\bibfnamefont
  {F.}~\bibnamefont {Bert}}, \bibinfo {author} {\bibfnamefont {P.}~\bibnamefont
  {Mendels}}, \bibinfo {author} {\bibfnamefont {R.~H.}\ \bibnamefont {Colman}},
  \ and\ \bibinfo {author} {\bibfnamefont {A.~S.}\ \bibnamefont {Wills}},\
  }\href {\doibase 10.1103/PhysRevB.87.155107} {\bibfield  {journal} {\bibinfo
  {journal} {Phys. Rev. B}\ }\textbf {\bibinfo {volume} {87}},\ \bibinfo
  {pages} {155107} (\bibinfo {year} {2013})}\BibitemShut {NoStop}%
\bibitem [{\citenamefont {Kermarrec}\ \emph {et~al.}(2014)\citenamefont
  {Kermarrec}, \citenamefont {Zorko}, \citenamefont {Bert}, \citenamefont
  {Colman}, \citenamefont {Koteswararao}, \citenamefont {Bouquet},
  \citenamefont {Bonville}, \citenamefont {Hillier}, \citenamefont {Amato},
  \citenamefont {van Tol}, \citenamefont {Ozarowski}, \citenamefont {Wills},\
  and\ \citenamefont {Mendels}}]{Kermarrec-2014}%
  \BibitemOpen
  \bibfield  {author} {\bibinfo {author} {\bibfnamefont {E.}~\bibnamefont
  {Kermarrec}}, \bibinfo {author} {\bibfnamefont {A.}~\bibnamefont {Zorko}},
  \bibinfo {author} {\bibfnamefont {F.}~\bibnamefont {Bert}}, \bibinfo {author}
  {\bibfnamefont {R.~H.}\ \bibnamefont {Colman}}, \bibinfo {author}
  {\bibfnamefont {B.}~\bibnamefont {Koteswararao}}, \bibinfo {author}
  {\bibfnamefont {F.}~\bibnamefont {Bouquet}}, \bibinfo {author} {\bibfnamefont
  {P.}~\bibnamefont {Bonville}}, \bibinfo {author} {\bibfnamefont
  {A.}~\bibnamefont {Hillier}}, \bibinfo {author} {\bibfnamefont
  {A.}~\bibnamefont {Amato}}, \bibinfo {author} {\bibfnamefont
  {J.}~\bibnamefont {van Tol}}, \bibinfo {author} {\bibfnamefont
  {A.}~\bibnamefont {Ozarowski}}, \bibinfo {author} {\bibfnamefont {A.~S.}\
  \bibnamefont {Wills}}, \ and\ \bibinfo {author} {\bibfnamefont
  {P.}~\bibnamefont {Mendels}},\ }\href {\doibase 10.1103/PhysRevB.90.205103}
  {\bibfield  {journal} {\bibinfo  {journal} {Phys. Rev. B}\ }\textbf {\bibinfo
  {volume} {90}},\ \bibinfo {pages} {205103} (\bibinfo {year}
  {2014})}\BibitemShut {NoStop}%
\bibitem [{\citenamefont {Reuther}\ \emph
  {et~al.}(2011{\natexlab{a}})\citenamefont {Reuther}, \citenamefont {Abanin},\
  and\ \citenamefont {Thomale}}]{Reuther-2011b}%
  \BibitemOpen
  \bibfield  {author} {\bibinfo {author} {\bibfnamefont {J.}~\bibnamefont
  {Reuther}}, \bibinfo {author} {\bibfnamefont {D.~A.}\ \bibnamefont {Abanin}},
  \ and\ \bibinfo {author} {\bibfnamefont {R.}~\bibnamefont {Thomale}},\ }\href
  {\doibase 10.1103/PhysRevB.84.014417} {\bibfield  {journal} {\bibinfo
  {journal} {Phys. Rev. B}\ }\textbf {\bibinfo {volume} {84}},\ \bibinfo
  {pages} {014417} (\bibinfo {year} {2011}{\natexlab{a}})}\BibitemShut
  {NoStop}%
\bibitem [{\citenamefont {Reuther}\ \emph
  {et~al.}(2011{\natexlab{b}})\citenamefont {Reuther}, \citenamefont
  {Thomale},\ and\ \citenamefont {Trebst}}]{Reuther-2011c}%
  \BibitemOpen
  \bibfield  {author} {\bibinfo {author} {\bibfnamefont {J.}~\bibnamefont
  {Reuther}}, \bibinfo {author} {\bibfnamefont {R.}~\bibnamefont {Thomale}}, \
  and\ \bibinfo {author} {\bibfnamefont {S.}~\bibnamefont {Trebst}},\ }\href
  {\doibase 10.1103/PhysRevB.84.100406} {\bibfield  {journal} {\bibinfo
  {journal} {Phys. Rev. B}\ }\textbf {\bibinfo {volume} {84}},\ \bibinfo
  {pages} {100406} (\bibinfo {year} {2011}{\natexlab{b}})}\BibitemShut
  {NoStop}%
\bibitem [{\citenamefont {Metzner}\ \emph {et~al.}(2012)\citenamefont
  {Metzner}, \citenamefont {Salmhofer}, \citenamefont {Honerkamp},
  \citenamefont {Meden},\ and\ \citenamefont {Sch\"onhammer}}]{Metzner-2012}%
  \BibitemOpen
  \bibfield  {author} {\bibinfo {author} {\bibfnamefont {W.}~\bibnamefont
  {Metzner}}, \bibinfo {author} {\bibfnamefont {M.}~\bibnamefont {Salmhofer}},
  \bibinfo {author} {\bibfnamefont {C.}~\bibnamefont {Honerkamp}}, \bibinfo
  {author} {\bibfnamefont {V.}~\bibnamefont {Meden}}, \ and\ \bibinfo {author}
  {\bibfnamefont {K.}~\bibnamefont {Sch\"onhammer}},\ }\href {\doibase
  10.1103/RevModPhys.84.299} {\bibfield  {journal} {\bibinfo  {journal} {Rev.
  Mod. Phys.}\ }\textbf {\bibinfo {volume} {84}},\ \bibinfo {pages} {299}
  (\bibinfo {year} {2012})}\BibitemShut {NoStop}%
\bibitem [{\citenamefont {Platt}\ \emph {et~al.}(2013)\citenamefont {Platt},
  \citenamefont {Hanke},\ and\ \citenamefont {Thomale}}]{Platt-2013}%
  \BibitemOpen
  \bibfield  {author} {\bibinfo {author} {\bibfnamefont {C.}~\bibnamefont
  {Platt}}, \bibinfo {author} {\bibfnamefont {W.}~\bibnamefont {Hanke}}, \ and\
  \bibinfo {author} {\bibfnamefont {R.}~\bibnamefont {Thomale}},\ }\href
  {\doibase 10.1080/00018732.2013.862020} {\bibfield  {journal} {\bibinfo
  {journal} {Adv. Phys.}\ }\textbf {\bibinfo {volume} {62}},\ \bibinfo {pages}
  {453} (\bibinfo {year} {2013})}\BibitemShut {NoStop}%
\bibitem [{\citenamefont {Katanin}(2004)}]{Katanin-2004}%
  \BibitemOpen
  \bibfield  {author} {\bibinfo {author} {\bibfnamefont {A.~A.}\ \bibnamefont
  {Katanin}},\ }\href {\doibase 10.1103/PhysRevB.70.115109} {\bibfield
  {journal} {\bibinfo  {journal} {Phys. Rev. B}\ }\textbf {\bibinfo {volume}
  {70}},\ \bibinfo {pages} {115109} (\bibinfo {year} {2004})}\BibitemShut
  {NoStop}%
\bibitem [{\citenamefont {Albuquerque}\ \emph {et~al.}(2011)\citenamefont
  {Albuquerque}, \citenamefont {Schwandt}, \citenamefont {Het\'enyi},
  \citenamefont {Capponi}, \citenamefont {Mambrini},\ and\ \citenamefont
  {L\"auchli}}]{Albuquerque-2011}%
  \BibitemOpen
  \bibfield  {author} {\bibinfo {author} {\bibfnamefont {A.~F.}\ \bibnamefont
  {Albuquerque}}, \bibinfo {author} {\bibfnamefont {D.}~\bibnamefont
  {Schwandt}}, \bibinfo {author} {\bibfnamefont {B.}~\bibnamefont {Het\'enyi}},
  \bibinfo {author} {\bibfnamefont {S.}~\bibnamefont {Capponi}}, \bibinfo
  {author} {\bibfnamefont {M.}~\bibnamefont {Mambrini}}, \ and\ \bibinfo
  {author} {\bibfnamefont {A.~M.}\ \bibnamefont {L\"auchli}},\ }\href {\doibase
  10.1103/PhysRevB.84.024406} {\bibfield  {journal} {\bibinfo  {journal} {Phys.
  Rev. B}\ }\textbf {\bibinfo {volume} {84}},\ \bibinfo {pages} {024406}
  (\bibinfo {year} {2011})}\BibitemShut {NoStop}%
\bibitem [{\citenamefont {Chaloupka}\ \emph {et~al.}(2010)\citenamefont
  {Chaloupka}, \citenamefont {Jackeli},\ and\ \citenamefont
  {Khaliullin}}]{Chaloupka-2010}%
  \BibitemOpen
  \bibfield  {author} {\bibinfo {author} {\bibfnamefont {J.}~\bibnamefont
  {Chaloupka}}, \bibinfo {author} {\bibfnamefont {G.}~\bibnamefont {Jackeli}},
  \ and\ \bibinfo {author} {\bibfnamefont {G.}~\bibnamefont {Khaliullin}},\
  }\href {\doibase 10.1103/PhysRevLett.105.027204} {\bibfield  {journal}
  {\bibinfo  {journal} {Phys. Rev. Lett.}\ }\textbf {\bibinfo {volume} {105}},\
  \bibinfo {pages} {027204} (\bibinfo {year} {2010})}\BibitemShut {NoStop}%
\bibitem [{\citenamefont {Perdew}\ \emph {et~al.}(1996)\citenamefont {Perdew},
  \citenamefont {Burke},\ and\ \citenamefont {Ernzerhof}}]{GGA}%
  \BibitemOpen
  \bibfield  {author} {\bibinfo {author} {\bibfnamefont {J.~P.}\ \bibnamefont
  {Perdew}}, \bibinfo {author} {\bibfnamefont {K.}~\bibnamefont {Burke}}, \
  and\ \bibinfo {author} {\bibfnamefont {M.}~\bibnamefont {Ernzerhof}},\ }\href
  {\doibase 10.1103/PhysRevLett.77.3865} {\bibfield  {journal} {\bibinfo
  {journal} {Phys. Rev. Lett.}\ }\textbf {\bibinfo {volume} {77}},\ \bibinfo
  {pages} {3865} (\bibinfo {year} {1996})}\BibitemShut {NoStop}%
\bibitem [{\citenamefont {Kresse}\ and\ \citenamefont
  {Hafner}(1993)}]{Kresse-1993}%
  \BibitemOpen
  \bibfield  {author} {\bibinfo {author} {\bibfnamefont {G.}~\bibnamefont
  {Kresse}}\ and\ \bibinfo {author} {\bibfnamefont {J.}~\bibnamefont
  {Hafner}},\ }\href {\doibase 10.1103/PhysRevB.47.558} {\bibfield  {journal}
  {\bibinfo  {journal} {Phys. Rev. B}\ }\textbf {\bibinfo {volume} {47}},\
  \bibinfo {pages} {558} (\bibinfo {year} {1993})}\BibitemShut {NoStop}%
\bibitem [{\citenamefont {Kresse}\ and\ \citenamefont
  {Furthmuller}(1996)}]{Kresse-1996}%
  \BibitemOpen
  \bibfield  {author} {\bibinfo {author} {\bibfnamefont {G.}~\bibnamefont
  {Kresse}}\ and\ \bibinfo {author} {\bibfnamefont {J.}~\bibnamefont
  {Furthmuller}},\ }\href {\doibase
  http://dx.doi.org/10.1016/0927-0256(96)00008-0} {\bibfield  {journal}
  {\bibinfo  {journal} {Computational Materials Science}\ }\textbf {\bibinfo
  {volume} {6}},\ \bibinfo {pages} {15 } (\bibinfo {year} {1996})}\BibitemShut
  {NoStop}%
\bibitem [{\citenamefont {Koepernik}\ and\ \citenamefont
  {Eschrig}(1999)}]{Koepernik-1999}%
  \BibitemOpen
  \bibfield  {author} {\bibinfo {author} {\bibfnamefont {K.}~\bibnamefont
  {Koepernik}}\ and\ \bibinfo {author} {\bibfnamefont {H.}~\bibnamefont
  {Eschrig}},\ }\href {\doibase 10.1103/PhysRevB.59.1743} {\bibfield  {journal}
  {\bibinfo  {journal} {Phys. Rev. B}\ }\textbf {\bibinfo {volume} {59}},\
  \bibinfo {pages} {1743} (\bibinfo {year} {1999})}\BibitemShut {NoStop}%
\bibitem [{FPL()}]{FPLO-webpage}%
  \BibitemOpen
  \href {http://www.FPLO.de} {\bibinfo  {journal} {www.FPLO.de}\ }\BibitemShut
  {NoStop}%
\bibitem [{\citenamefont {Messio}\ \emph {et~al.}(2011)\citenamefont {Messio},
  \citenamefont {Lhuillier},\ and\ \citenamefont {Misguich}}]{Messio-2011}%
  \BibitemOpen
\bibfield  {journal} {  }\bibfield  {author} {\bibinfo {author} {\bibfnamefont
  {L.}~\bibnamefont {Messio}}, \bibinfo {author} {\bibfnamefont
  {C.}~\bibnamefont {Lhuillier}}, \ and\ \bibinfo {author} {\bibfnamefont
  {G.}~\bibnamefont {Misguich}},\ }\href {\doibase 10.1103/PhysRevB.83.184401}
  {\bibfield  {journal} {\bibinfo  {journal} {Phys. Rev. B}\ }\textbf {\bibinfo
  {volume} {83}},\ \bibinfo {pages} {184401} (\bibinfo {year}
  {2011})}\BibitemShut {NoStop}%
\bibitem [{\citenamefont {Gong}\ \emph {et~al.}(2015)\citenamefont {Gong},
  \citenamefont {Zhu}, \citenamefont {Balents},\ and\ \citenamefont
  {Sheng}}]{Gong-2015}%
  \BibitemOpen
  \bibfield  {author} {\bibinfo {author} {\bibfnamefont {S.-S.}\ \bibnamefont
  {Gong}}, \bibinfo {author} {\bibfnamefont {W.}~\bibnamefont {Zhu}}, \bibinfo
  {author} {\bibfnamefont {L.}~\bibnamefont {Balents}}, \ and\ \bibinfo
  {author} {\bibfnamefont {D.~N.}\ \bibnamefont {Sheng}},\ }\href {\doibase
  10.1103/PhysRevB.91.075112} {\bibfield  {journal} {\bibinfo  {journal} {Phys.
  Rev. B}\ }\textbf {\bibinfo {volume} {91}},\ \bibinfo {pages} {075112}
  (\bibinfo {year} {2015})}\BibitemShut {NoStop}%
\bibitem [{\citenamefont {Bieri}\ \emph {et~al.}(2015)\citenamefont {Bieri},
  \citenamefont {Messio}, \citenamefont {Bernu},\ and\ \citenamefont
  {Lhuillier}}]{Bieri-2015}%
  \BibitemOpen
  \bibfield  {author} {\bibinfo {author} {\bibfnamefont {S.}~\bibnamefont
  {Bieri}}, \bibinfo {author} {\bibfnamefont {L.}~\bibnamefont {Messio}},
  \bibinfo {author} {\bibfnamefont {B.}~\bibnamefont {Bernu}}, \ and\ \bibinfo
  {author} {\bibfnamefont {C.}~\bibnamefont {Lhuillier}},\ }\href {\doibase
  10.1103/PhysRevB.92.060407} {\bibfield  {journal} {\bibinfo  {journal} {Phys.
  Rev. B}\ }\textbf {\bibinfo {volume} {92}},\ \bibinfo {pages} {060407}
  (\bibinfo {year} {2015})}\BibitemShut {NoStop}%
\bibitem [{\citenamefont {F\aa{}k}\ \emph {et~al.}(2012)\citenamefont
  {F\aa{}k}, \citenamefont {Kermarrec}, \citenamefont {Messio}, \citenamefont
  {Bernu}, \citenamefont {Lhuillier}, \citenamefont {Bert}, \citenamefont
  {Mendels}, \citenamefont {Koteswararao}, \citenamefont {Bouquet},
  \citenamefont {Ollivier}, \citenamefont {Hillier}, \citenamefont {Amato},
  \citenamefont {Colman},\ and\ \citenamefont {Wills}}]{Fak-2012}%
  \BibitemOpen
  \bibfield  {author} {\bibinfo {author} {\bibfnamefont {B.}~\bibnamefont
  {F\aa{}k}}, \bibinfo {author} {\bibfnamefont {E.}~\bibnamefont {Kermarrec}},
  \bibinfo {author} {\bibfnamefont {L.}~\bibnamefont {Messio}}, \bibinfo
  {author} {\bibfnamefont {B.}~\bibnamefont {Bernu}}, \bibinfo {author}
  {\bibfnamefont {C.}~\bibnamefont {Lhuillier}}, \bibinfo {author}
  {\bibfnamefont {F.}~\bibnamefont {Bert}}, \bibinfo {author} {\bibfnamefont
  {P.}~\bibnamefont {Mendels}}, \bibinfo {author} {\bibfnamefont
  {B.}~\bibnamefont {Koteswararao}}, \bibinfo {author} {\bibfnamefont
  {F.}~\bibnamefont {Bouquet}}, \bibinfo {author} {\bibfnamefont
  {J.}~\bibnamefont {Ollivier}}, \bibinfo {author} {\bibfnamefont {A.~D.}\
  \bibnamefont {Hillier}}, \bibinfo {author} {\bibfnamefont {A.}~\bibnamefont
  {Amato}}, \bibinfo {author} {\bibfnamefont {R.~H.}\ \bibnamefont {Colman}}, \
  and\ \bibinfo {author} {\bibfnamefont {A.~S.}\ \bibnamefont {Wills}},\ }\href
  {\doibase 10.1103/PhysRevLett.109.037208} {\bibfield  {journal} {\bibinfo
  {journal} {Phys. Rev. Lett.}\ }\textbf {\bibinfo {volume} {109}},\ \bibinfo
  {pages} {037208} (\bibinfo {year} {2012})}\BibitemShut {NoStop}%
\bibitem [{\citenamefont {Janson}\ \emph {et~al.}(2008)\citenamefont {Janson},
  \citenamefont {Richter},\ and\ \citenamefont {Rosner}}]{Janson-2008}%
  \BibitemOpen
  \bibfield  {author} {\bibinfo {author} {\bibfnamefont {O.}~\bibnamefont
  {Janson}}, \bibinfo {author} {\bibfnamefont {J.}~\bibnamefont {Richter}}, \
  and\ \bibinfo {author} {\bibfnamefont {H.}~\bibnamefont {Rosner}},\ }\href
  {\doibase 10.1103/PhysRevLett.101.106403} {\bibfield  {journal} {\bibinfo
  {journal} {Phys. Rev. Lett.}\ }\textbf {\bibinfo {volume} {101}},\ \bibinfo
  {pages} {106403} (\bibinfo {year} {2008})}\BibitemShut {NoStop}%
\end{thebibliography}
\end{document}